\documentclass[journal=cmatex,manuscript=article,layout=twocolumn]{achemso}

\usepackage{array}
\usepackage{textcomp}
\usepackage{gensymb}
\usepackage{graphicx}
\usepackage{multirow}
\usepackage[super]{nth}

\usepackage{amsmath}
\newcommand\ce[1]{\ensuremath{\mathrm{#1}}}
\usepackage{xcolor} 

\setkeys{acs}{usetitle=true} 
\SectionNumbersOn

\let\oldmaketitle\maketitle
\let\maketitle\relax

\title{On the Active Components in Crystalline Li-Nb-O and Li-Ta-O Coatings from First Principles}

\author{Hengning Chen}
\affiliation{Department of Materials Science and Engineering, National University of Singapore, 9 Engineering Drive 1, 117575, Singapore}

\author{Zeyu Deng}
\affiliation{Department of Materials Science and Engineering, National University of Singapore, 9 Engineering Drive 1, 117575, Singapore}

\author{Yuheng Li}
\affiliation{Department of Materials Science and Engineering, National University of Singapore, 9 Engineering Drive 1, 117575, Singapore}

\author{Pieremanuele Canepa}
\affiliation{Department of Materials Science and Engineering, National University of Singapore, 9 Engineering Drive 1, 117575, Singapore}
\altaffiliation{Department of Chemical and Biomolecular Engineering, National University of Singapore, 4 Engineering Drive 4, 117585, Singapore}
\email{pcanepa@nus.edu.sg}

\begin{document}


\twocolumn[
\begin{@twocolumnfalse}
\oldmaketitle
\begin{abstract}
\noindent Layered-oxide \ce{LiNi_xMn_yCo_{1-x-y}O_2} (NMC) positive electrodes with high Nickel content, deliver high voltages and energy densities. However, a high nickel content, e.g., x = 0.8 (NMC 811), can lead to high surface reactivity, which can trigger thermal runaway and gas generation. While claimed safer, all-solid-state batteries still suffer from high interfacial resistance. Here, we investigate niobate and tantalate coating materials, which can mitigate the interfacial reactivities in Li-ion and all-solid-state batteries. First-principles calculations reveal the multiphasic nature of Li--Nb--O and Li--Ta--O coatings, containing mixtures of {\ce{LiNbO_3}} and {\ce{Li_3NbO_4}}, or of {\ce{LiTaO_3}} and {\ce{Li_3TaO_4}}. The concurrence of several phases in Li--Nb--O or Li--Ta--O modulates the type of stable native defects in these coatings. Li--Nb--O and Li--Ta--O coating materials can form favorably lithium vacancies {\ce{Vac^{'}_{Li}}} and antisite defects {\ce{Nb^{\bullet \bullet \bullet \bullet}_{Li}}} ({\ce{Ta^{\bullet \bullet \bullet \bullet}_{Li}}}) combined into charge-neutral defect complexes. Even in defective crystalline \ce{LiNbO_3} (or \ce{LiTaO_3}), we reveal poor Li-ion conduction properties. In contrast, \ce{Li_3NbO_4} and \ce{Li_3TaO_4} that are introduced by high-temperature calcinations can provide adequate Li-ion transport in these coatings. Our in-depth investigation of the structure-property relationships in the important Li--Nb--O and Li--Ta--O coating materials helps to develop more suitable calcination protocols to maximize the functional properties of these niobates and tantalates. 
\end{abstract}
\end{@twocolumnfalse}
]




\section{Introduction}

Notwithstanding the maturity of the lithium (Li)-ion battery technology, stabilizing the electrode-electrolyte interfaces and interphases set challenges in lithium-ion batteries (LIBs), as well as in novel all-solid-state batteries (ASSBs).\cite{LiLiansheng2021LNcoatingreview, Wang2015interfacestabilityLIBs}  For example, in LIBs, the hydrofluoric acid released upon \ce{LiPF_6} decomposition can lead to the surface corrosion of cathode particles and the subsequent leaching of transition metal ions (TMs) into the electrolyte.\cite{Wang2015interfacestabilityLIBs} Furthermore, during reversible Li intercalation in cathode active materials (CAMs), significant volume variations of the active particles may induce cracks that lead to loss of contact among the active particles, and a subsequent reduction in performance.\cite{WenboZhang2017expansion} 

Thiophosphate solid electrolytes (SEs), e.g., \ce{Li_{10}GeP_2S_{12}} (LGPS) and argyrodites \ce{Li_6PS_5X} (with \ce{X}~=~\ce{Cl}, \ce{Br}, \ce{I})\cite{Piero2019SSEnature} display high Li-ion conductivities and low-temperature processability, and hence are promising candidates for ASSBs. However, the low oxidative (or anodic) stabilities of thiophosphate SEs can result in undesired side reactions, especially with high-voltage ($>$~4.5 V vs. \ce{Li}/\ce{Li^+}) CAMs.\cite{Tan2019thiophosphatestability,YizhouZhu2015thermodynamicSE} 

Introducing coating materials is one of the most effective solutions to mitigate possible interfacial instabilities in LIBs and ASSBs, improving their capacity retention, rate capabilities, and cells’ longevity.\cite{Minnmann2022SEcoating, ZonghaiChen2010coatingLIBs,Culver2019Funcionalitycoating} Coating materials can enhance the ``adhesion'' between cathode particles and act as physical protections from undesired chemical reactions. Coating materials also stabilize the cathode-electrolyte interfaces chemically and electrochemically, thereby increasing or modulating  electrolytes' stability windows. 

Simultaneously, coatings in LIBs are designed to have high ionic and electronic conductivities ---both properties are essential for maintaining percolating ionic and electronic networks between  cathode particles.\cite{Wang2015interfacestabilityLIBs} The requirements of coatings (for cathode electrodes) in ASSBs are different from LIBs: ({\emph{i}}) deliver high Li-ion conductivity, and ({\emph{ii}}) maintain sufficiently low electronic conductivity to prevent SEs with narrow stability windows from being easily oxidized by high-voltage CAMs.\cite{Culver2019Funcionalitycoating}

In particular, amorphous {\ce{LiNbO_3}} and {\ce{LiTaO_3}} materials with high ionic conductivities ($\sim$10$^{-5}$--10$^{-6}$~$\mathrm{S ~cm^{-1}}$ at room temperature) and appreciable electronic resistance ($\sim$10$^{-11}$--10$^{-12}$~$\mathrm{\Omega^{-1}~cm^{-1}}$) have been effectively used as coatings in ASSBs.\cite{NarumiOhta2007,SSBNCM712LNO2021Peng,GaozhanLiu2020LNcoating5VSSB} Albeit with much lower Li-ion conductivities ($<$10$^{-12}$~$\mathrm{S ~cm^{-1}}$ at 400K) and exceedingly high migration barriers (1.16--1.33~eV),\cite{Heitjans2006NMRsingle, Wilkening2008LTnanocrystalline} crystalline {\ce{LiNbO_3}} and {\ce{LiTaO_3}} coatings show substantial enhancements in the electrochemical performance of LIBs (see Table S1 of the Supporting Information, SI).\cite{YantaoZhang2014,GuozhongLu2021Study,FengxiaXin2021WhatistheRole,JunSuLee2021Comparison} 

As the information available in the literature on amorphous and crystalline phases of niobates and tantalates coatings appears scattered, it is important to set a firm baseline on the crystalline phases of these materials. While amorphous niobates and tantalates appear to provide the highest ionic conductivities,{\cite{Glass1978Ionic,NarumiOhta2007,Heitjans2006NMRsingle}} the crystalline analogues are reported to increase the hardness of cathode particles.{\cite{Kim2020Bifunctional,HyoBinLee2022Surface}}

Although higher levels of crystallinity in {\ce{LiNbO_3}} (or {\ce{LiTaO_3}}) have been linked to a reduction of CAMs particle cracking,\cite{Kim2020Bifunctional, HyoBinLee2022Surface} there exists an apparent contradiction between the low Li-ion conductivity of crystalline {\ce{LiNbO_3}} (or {\ce{LiTaO_3}}) and the good rate capabilities achieved in these cells.\cite{Glass1978Ionic,YantaoZhang2014} Indeed, the texture and composition of these niobate and tantalate materials appear evidently more complicated than commonly perceived. 

Previous reports{\cite{Wilkening2008LTnanocrystalline,Heitjans2006NMRsingle}} suggested that highly defective nanocrystalline \ce{LiNbO_3} and \ce{LiTaO_3} (from high-energy ball milling) containing amorphous-like components exhibit room-temperature Li-ion conductivities ($\sim$10$^{-6}$~$\mathrm{S~cm^{-1}}$), comparable with their amorphous analogs (10$^{-5}$--10$^{-6}$~$\mathrm{S ~cm^{-1}}$). These facts hint at the high sensitivity of the Li-ion conductivity in {\ce{LiNbO_3}} and {\ce{LiTaO_3}} to the presence of defects in their crystalline phases. 

Conventional synthesis protocols of crystalline niobate (or tantalate) coatings typically apply the wet-chemical method followed by prolonged (3 to 10 hour-long) calcinations at 773--1073K (Table~S1). Furthermore, {\ce{LiNbO_3}} (or {\ce{LiTaO_3}}) tend to crystallize out from the congruent melts (Li--poor, M--rich) at off-stochiometric compositions facilitating the incorporation of the intrinsic defects (see Figure~\ref{tieline_pd}).{\cite{Iyi1992comparative,Smyth1983LNdefect,Kolb1976pdLTO,Bernasconi1999LTantisite}} However, neutron reflectometry and secondary ion mass spectroscopy measurements did not suggest a marked increase in Li-ion transport in these materials.{\cite{Rahn2011congruentLNconductivity,Huger2014singleLiNbO3Ea}} Therefore, the role played by intrinsic defects in the functional properties of niobate and tantalate coatings remains ambiguous.

Recent investigations on the composition of Li--Nb--O coating highlight the coexistence of two primary phases: \ce{LiNbO_3} and \ce{Li_3NbO_4}.\cite{YantaoZhang2014,GuozhongLu2021Study,Xin2021Conditioning} \citeauthor{YantaoZhang2014}\cite{YantaoZhang2014} suggested that the coexistence of \ce{LiNbO_3} and \ce{Li_3NbO_4} improve the electrochemical performances of coated cathodes. Several reports suggested that the {\ce{Li^+}}-migration barrier in crystalline \ce{Li_3NbO_4} (0.58--0.85~eV) is considerably lower than that of crystalline \ce{LiNbO_3} (1.16--1.33~eV),\cite{Glass1978Ionic,Liao2013L3NO4barrierWsubstitution} \citeauthor{Liao2013L3NO4barrierWsubstitution} and \citeauthor{Yabuuchi2015Li3NbO4electrode} showed an increase in Li-ion conductivity by several orders of magnitude in {\ce{Li_3NbO_4}} upon doping with \ce{W^{6+}}, or substituting of \ce{Ni^{2+}} for \ce{Li^+}, which introduces Li vacancies.\cite{Liao2013L3NO4barrierWsubstitution,Yabuuchi2015Li3NbO4electrode} This evidence suggests {\ce{Li_3NbO_4}} may be the active phases in transporting Li-ions in niobate coatings. 

Based on the facts exposed, protocols employed for synthesizing crystalline niobate coatings can strongly affect: (\emph{i}) the partition of the active phases of the Li--Nb--O coatings,\cite{Xin2021Conditioning, GuozhongLu2021Study} (\emph{ii}) the concentrations of intrinsic defects, and in turn affect (\emph{iii}) the Li-ion diffusion in these materials. This knowledge gap requires immediate attention to optimize the functional properties of niobates and tantalates.

Leveraging a combination of first-principles calculations (density functional theory), bond valence mapping (through the \emph{SoftBV}),{\cite{adams2006bond}} nudged-elastic band (NEB) calculations, and machine learning (ML) molecular dynamics (MDs), we investigate how the Li-migration properties of {\ce{LiMO_3}} and {\ce{Li_3MO_4}} (with {\ce{M}}~=~{\ce{Nb}} and {\ce{Ta}}) are modulated by the occurrence of thermodynamically viable intrinsic defects.  We demonstrate that what is commonly perceived as a {\ce{LiNbO_3}} (or {\ce{LiTaO_3}}) coating material, in practice is a multiphasic composite including variable fractions of {\ce{Li_3NbO_4}} (or {\ce{Li_3TaO_4}}), altering the conduction property of the material.   The occurrence of intrinsic defects in niobates (and tantalates) remains conditional to the coexistence of \ce{LiMO_3} and \ce{Li_3MO_4} at synthesis conditions, which regulates the availability of Li vacancies. A progressive transformation of {\ce{LiMO_3}} into {\ce{Li_3MO_4}} by prolonged high-temperature calcination results in a much less defective \ce{Li_3MO_4} coating material, which in turn hinders the \ce{Li^+} transport and deteriorates the functionality of the coating.

\section{Crystal Structures of \ce{\mathbf{LiMO_3}} and \ce{\mathbf{Li_3MO_4}}} 
\label{structure}
\begin{figure*}[!ht]
    \centering
    \includegraphics[width=0.9\textwidth]{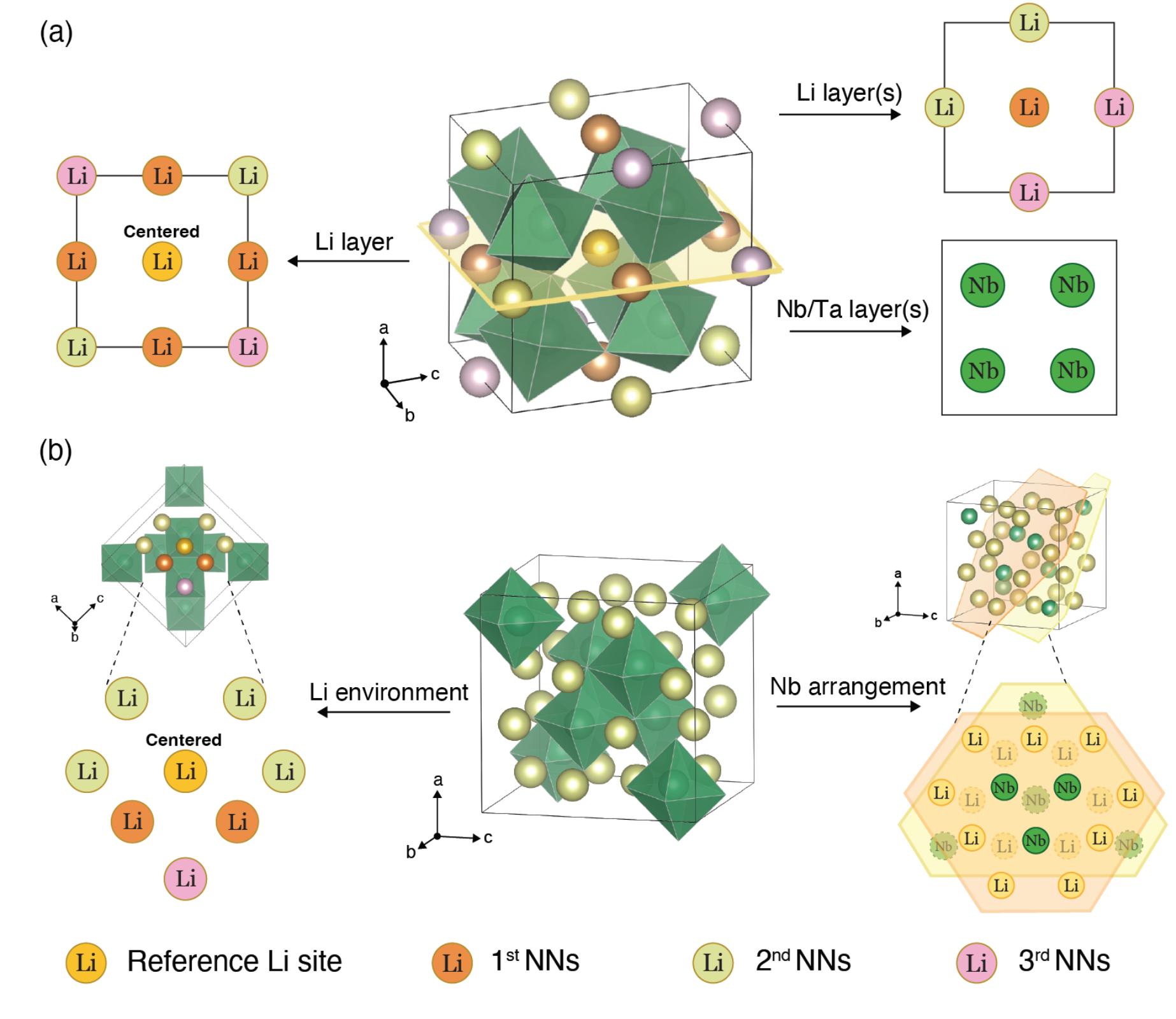}
    \caption{
    \label{structure_LNL3N}
    (a) Structures of {\ce{LiNbO_3}} ({\ce{LiTaO_3}}) ($R3c$) with corner-sharing \ce{Nb(Ta)O_6} octahedra, transformed into pseudo-cubic structures using $\left[[1, 1, -1],\ [-1, 1, 1],\ [1, -1, 1]\right]$ matrix for visualization. (b) Structure of {\ce{Li_3NbO_4}} ($I\overline{4}3m$) with edge-sharing \ce{Nb(Ta)O_6} octahedra. For the centered Li site in {\ce{LiNbO_3}} ({\ce{LiTaO_3}}) in (a), there are two \nth{1} near neighbor (NNs), four \nth{2} NNs, four \nth{3} NNs at the top and at the bottom Li layers, and four \nth{1} NNs, two \nth{2} NNs, two \nth{3} NNs on the middle Li layer, as well as eight nearby Nb(Ta) atoms. The left panel in (b) describes the local environment of a Li site in {\ce{Li_3NbO_4}}, and the right panel shows the arrangement of four Nb atoms within the {\ce{Nb_4O_{16}}} clusters at the center of orange and yellow planes. Additional views of these structures emphasizing the Li-ion environments are presented in Figure~S2 of the SI.
    }
\end{figure*}

 Figure~{\ref{structure_LNL3N}} shows the \ce{LiNbO_3} and \ce{LiTaO_3} perovskite structures (space group $R3c$), which present octahedra tilting.\cite{Abrahams1966LNR3c,Volk2008LNphasediagram}  In \ce{LiNbO_3} (\ce{LiTaO_3}), surrounded by oxygen atoms, Li and Nb (Ta) cations occupying 2/3 octahedral sites are aligned along the \textit{c} axis (see Figure S1 of the SI) and follow a stacking sequence of Nb (Ta), Li (or vacancy), Nb (Ta).\cite{Iyi1992comparative} Both \ce{LiNbO_3} and \ce{LiTaO_3} show a single Li Wyckoff site.
 
We analyze the immediate environment of Li ions (gold atom in Figure~\ref{structure_LNL3N}a), which is important to set up different models of intrinsic defect complexes and investigate Li-ion migration pathways in \ce{LiNbO_3} and \ce{LiTaO_3}. For a Li site in \ce{LiNbO_3} and \ce{LiTaO_3}, there are six \nth{1} nearest neighbors (NNs), six \nth{2} NNs, and six \nth{3} NNs Li sites in total from the top, middle and bottom Li layers, respectively. The eight neighboring \ce{Nb(Ta)O_6} are shown in Figure~\ref{structure_LNL3N}a as green octahedra.

\ce{Li_3NbO_4} shows a disordered rock-salt structure with an average $Fm\overline{3}m$ space group (Figure S3 of the SI) when synthesized at low temperature (at $\sim$220~\celsius).\cite{MayNyman2009L3NL3Tsynthesis, Grenier1964Fm3mLNLT} At 700~\celsius, \ce{Li_3NbO_4} transforms into an $I\overline{4}3m$ ordered phase  (Figure~\ref{structure_LNL3N}b),\cite{Hsiao2010L3Nbandgap,Deena2010Li3NO4disorderedtransform,Mather2000Li3NbO4structure} which highlights the edge-sharing \ce{NbO_6} octahedra. In the right panel of Figure~\ref{structure_LNL3N}b, three Nb cations (at the center) form a tetramer ({\ce{Nb_4O_{16}}}) with the other Nb cation in subsequent layers. Together with the Li ions (and vacancies), these \ce{Nb_4O_{16}} clusters form a body-centered cubic lattice. Compared with the $R3c$ closely-packed \ce{LiNbO_3}, this arrangement might allow larger voids that  facilitate Li transport. For each Li site, there are two \nth{1} NNs, four \nth{2} NNs, and one \nth{3} NNs as shown in the left panel of Figure~\ref{structure_LNL3N}b.

\begin{figure}[!ht]
    \centering
    \includegraphics[width=\columnwidth]{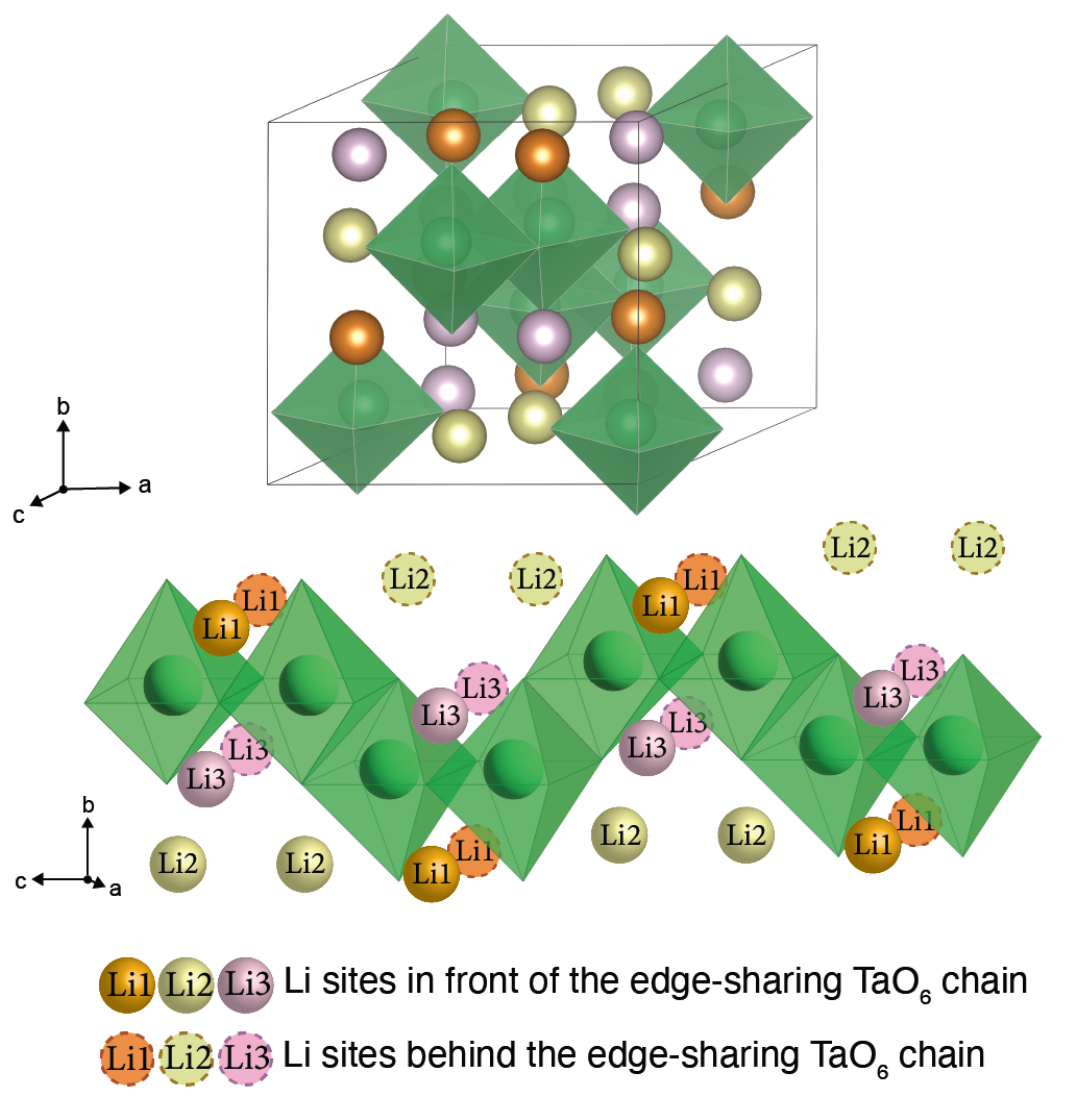}
    \caption{\label{structure_L3T}
    Structure of $\beta$-\ce{Li_3TaO_4} in the $C2/c$ space group. Unique Li sites are labeled and shown with different colors. The lower panel shows edge-sharing (green) \ce{TaO_6} octahedra zigzag along the $c$ direction. Li environments in $\beta$-{\ce{Li_3TaO_4}} are highlighted in Figure~S2c of the SI.
    }
\end{figure}

\ce{Li_3TaO_4} exists in three main polymorphs (Table S3): (\emph{i}) a disordered low-temperature phase with an average space group $Fm\overline{3}m$ (similar to {\ce{Li_3NbO_4}}, Figure S3), (\emph{ii}) an 
intermediate $\beta$-phase (with a space group $C2/c$), and (\emph{iii}) a high-temperature ($>$1450~\celsius) $\alpha$-phase structure (with a space group $P2$).\cite{MayNyman2009L3NL3Tsynthesis,Mather2000Li3NbO4structure,Zocchi1983Li3TaO4} Since the synthesis temperature (400-800 \celsius, Table~S1) of crystalline Li--Ta--O coating materials are usually lower than 1450~\celsius, the formation of the high-temperature $\alpha$-phase of \ce{Li_3TaO_4} can be excluded from the composition of typical Li--Ta--O coating materials. Therefore, we speculate that the ordered $\beta$-phase $C2/c$ \ce{Li_3TaO_4} phase can be the other constituent of the coating material together with the $R3c$ \ce{LiTaO_3} phase. The ordered $C2/c$ {\ce{Li_3TaO_4}} shows zigzag chains formed by the edge-sharing \ce{TaO_6} octahedra surrounded by Li ions (right panel of Figure \ref{structure_L3T}). \ce{Li_3TaO_4} contains eight independent crystallographic sites (three Li, one Ta, and four O) in the unit cell, \cite{Boulay2003betaL3Treinvestigation, ChaeeunKim2022Li3TaO4polymorphs} which denotes the substantial difference in terms of local arrangements compared to $R3c$ {\ce{LiNbO_3}}, {\ce{LiTaO_3}}, and $I{\overline{4}}3m$ {\ce{Li_3NbO_4}} displaying a unique Li site.

\section{Phase Equlibria and Electrochemical Stabilities of Li-Nb-O and Li-Ta-O Coating Materials} 
\label{electrochemical}

\begin{figure*}[!ht]
    \centering
    \includegraphics[width=1.0\textwidth]{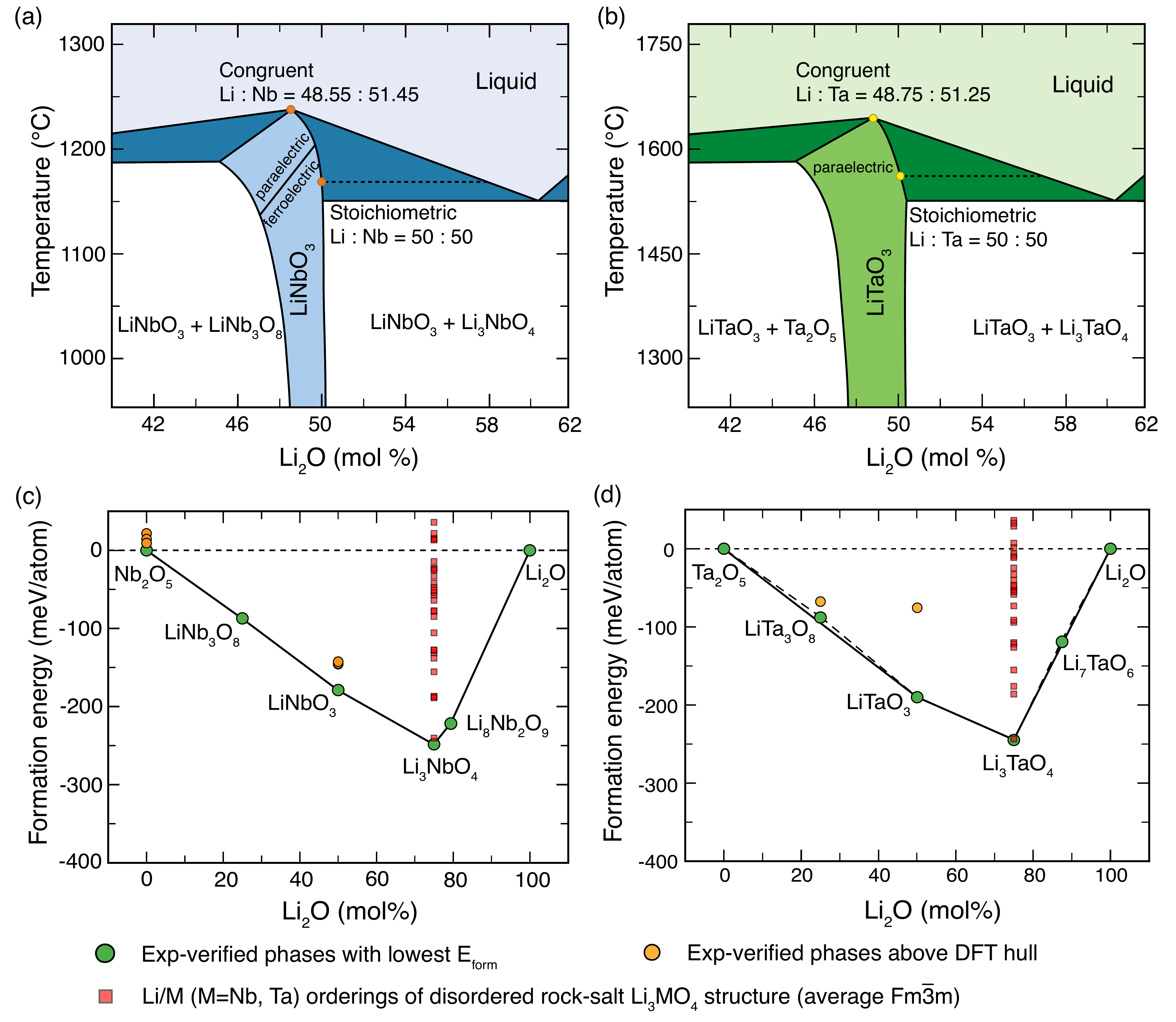}
    \caption{Phase diagrams of (a) \ce{Li_2O}--\ce{Nb_2O_5} and (b) \ce{Li_2O}--\ce{Ta_2O_5} pseudo-binary systems redrawn from previous experimental data.\cite{Volk2008LNphasediagram,Allemann1996LTphasediagram,Kolb1976pdLTO} The melting point of {\ce{LiNbO_3}} and \ce{LiTaO_3} is $\sim$1240~\celsius~and $\sim$1650~\celsius, respectively.{\cite{Volk2008LNphasediagram}} The  phase diagrams of (c) \ce{Li_2O}--\ce{Nb_2O_5}, and (d) \ce{Li_2O}--\ce{Ta_2O_5} computed from DFT (r$^2$SCAN) at 0K. Metastable or unstable Li/M orderings of {\ce{Li_3MO_4}} (M~=~Nb, Ta) are marked by red squares. The dashed line in (d) is a guide for the eye to show that {\ce{LiTa_3O_8}}  sits slightly above the stability line formed by {\ce{Ta_2O_5}} and {\ce{LiTaO_3}}.
    }
    \label{tieline_pd}
\end{figure*}

Figure~\ref{tieline_pd}a and Figure~\ref{tieline_pd}b show the experimental binary phase diagrams of the \ce{Li_2O}--\ce{Nb_2O_5} and \ce{Li_2O}--\ce{Ta_2O_5} systems in the composition range spanning from 40~mol\% to 62~mol\% of {\ce{Li_2O}}.\cite{Volk2008LNphasediagram,Allemann1996LTphasediagram,Kolb1976pdLTO} 

In Figure~\ref{tieline_pd}a, an exceeding (decreasing) \ce{Li_2O} content into the {\ce{LiNbO_3}} leads to the formation of a secondary phase, \ce{Li_3NbO_4} (\ce{LiNb_3O_8}), that coexists with \ce{LiNbO_3} over a large range of composition. In Figure~\ref{tieline_pd}b, \ce{Li_3TaO_4} appears as a secondary phase with \ce{LiTaO_3} in the \ce{Li_2O}-rich regions.

Figures~{\ref{tieline_pd}}c and {\ref{tieline_pd}}d show the computed (@r$^2$SCAN) phase diagrams at 0~K of the {\ce{Li_2O}}--{\ce{Nb_2O_5}} and {\ce{Li_2O}}--{\ce{Ta_2O_5}} pseudo-binary lines including all experimentally known compositions. Green circles show the lowest energy structures computed by DFT, while phases predicted as metastable by DFT but identified experimentally are shown by yellow circles. \ce{Li_8Ta_2O_9} has been excluded from Figure~\ref{tieline_pd}d due to the unavailability of accurate structural information. As for the disordered \ce{Li_3MO_4} (M~=~Nb, Ta), 32 unique orderings of Li/M have been generated for each composition using a $\sim$2$\times$2$\times$1 supercell model (including 4 f.u.). Red squares show the formation energies of each Li/M ordering.

Among all competing phases, the $R3c$ \ce{LiNbO_3} and the $I\overline{4}3m$ \ce{Li_3NbO_4} lie on the \ce{Nb_2O_5}-\ce{Li_2O} convex hull---the convex envelopes of structures with the lowest formation energies. Therefore, any Li-rich compositions of {\ce{LiNbO_3}} (addition of {\ce{Li_2O}} to {\ce{LiNbO_3}}) will be a mixture also containing {\ce{Li_3NbO_4}}, and in excellent agreement with Figure~\ref{tieline_pd}a. Similarly, the $R3c$ \ce{LiTaO_3} and the $C2/c$ \ce{Li_3TaO_4} are the most stable phases in the \ce{Ta_2O_5}--\ce{Li_2O} convex hull. Note, the formation energy of \ce{LiTa_3O_8} falls slightly above ($\sim$7 meV/atom) the convex hull set by \ce{Ta_2O_5} and \ce{LiTaO_3}. This is in agreement with the absence of \ce{LiTa_3O_8} phase as observed in experiments, where only \ce{Ta_2O_5} and \ce{LiTaO_3} phases form under the Li-poor conditions.\cite{Kolb1976pdLTO, Allemann1996LTphasediagram} The computed phase diagrams of Figure~\ref{tieline_pd} can qualitatively and quantitatively reproduce experimental observations.

\begin{figure*}[!ht]
    \centering
    \includegraphics[width=\textwidth]{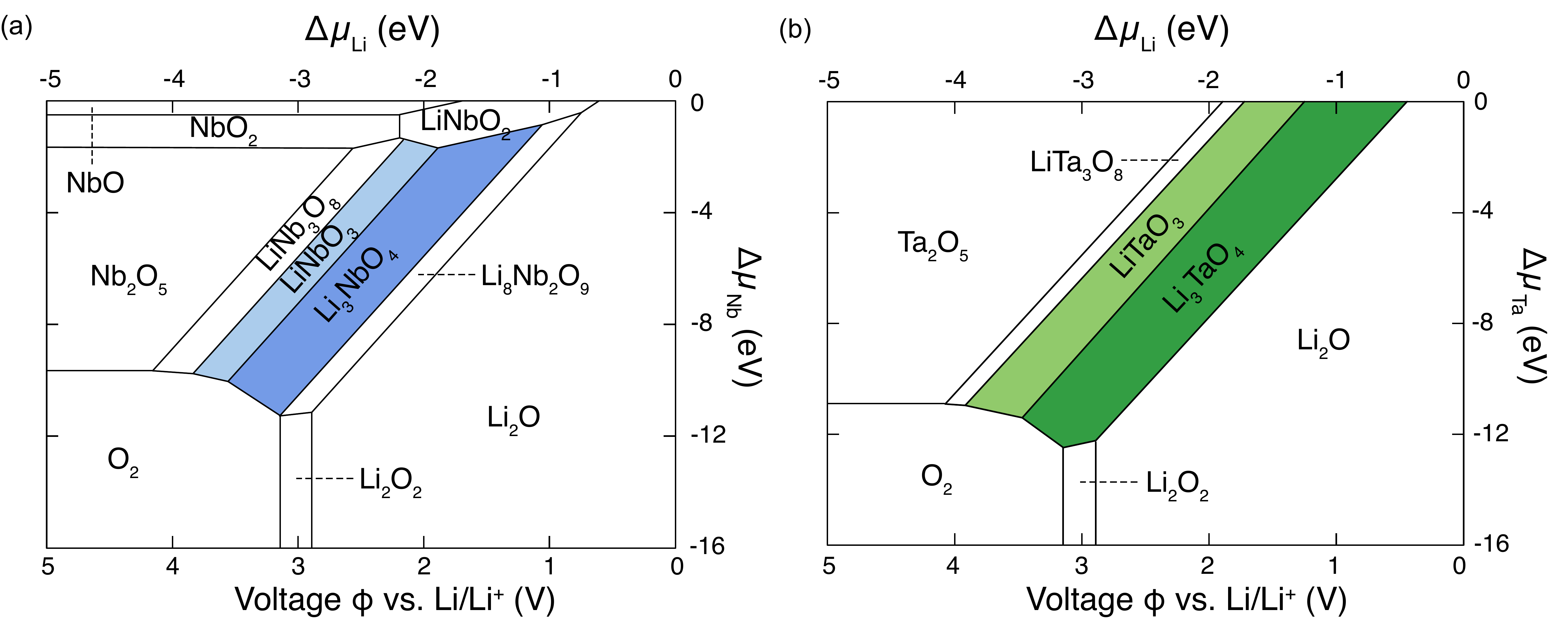}
    \caption{DFT-computed (with r$^2$SCAN and at 0 K) stability regions of niobate and tantalate phases in the (a) Li-Nb-O and (b) Li-Ta-O domains. Thermodynamically stable regions are labeled, of which \ce{LiNbO_3}, \ce{Li_3NbO_4}, \ce{LiTaO_3}, and \ce{Li_3TaO_4} are highlighted with different colors. Stable phases feature as a function of changes in chemical potential ($\Delta\mu_\mathrm{Li}$, $\Delta\mu_\mathrm{Nb}$ and $\Delta\mu_\mathrm{Ta}$) of Li, Nb (or Ta) referenced to their standard elemental states. Variations of $\Delta\mu_\mathrm{Li}$ can be easily converted into voltages (see main text). }
    \label{chempot_ref}
\end{figure*}

To evaluate the electrochemical stability windows of these coating materials, we construct the ternary Li--M--O (M~=~Nb, Ta) phase diagrams based on the elemental chemical potential changes, $\Delta\mu_\mathrm{Li}$ and $\Delta\mu_\mathrm{M}$, as seen in Figure~\ref{chempot_ref}. All competing phases available in the ICSD were included. 

The feasible ranges of $\Delta \mu_\mathrm{Li}$ and $\Delta \mu_\mathrm{M}$ in which the \ce{LiMO_3} and \ce{Li_3MO_4} are thermodynamically stable are shown by colored regions in Figure~\ref{chempot_ref}a and \ref{chempot_ref}b. $\Delta \mu_\mathrm{Li}$ is the Li chemical potential ($\mu_\mathrm{Li}$) referenced to the chemical potential of Li metal ($\mu_\mathrm{Li}^0$). $\mu_\mathrm{Li}$ is converted using the Nernst equation into voltages (versus Li/\ce{Li^+}), $\mathrm{\phi} = \frac{\mu_\mathrm{Li}^0 - \mu_\mathrm{Li}}{e\mathrm{F}} = -\frac{\Delta \mu_\mathrm{Li}}{e\mathrm{F}}$.

\begin{table*}[!ht]
\caption{Computed electrochemical stability windows (in V) of coating materials, and their reaction energies \ce{\Delta E_{rxt}} (meV/atom) with the different positive electrodes and solid electrolyte chemistries obtained from previous computational works.\cite{YihanXiao2019computationalcoatingscreen,YizhouZhu2016Firstprinciplecoating} Red. and Ox. are for reductive (anodic) stability and oxidative (cathodic) stability, respectively. NMC is the abbreviation for {\ce{LiNi_xMn_yCo_{1-x-y}O_2}}, LCO for {\ce{LiCoO_2}}, LMO for {\ce{Li_2Mn_2O_4}}, LFP for {\ce{LiFePO_4}}, and LLZO for {\ce{Li_7La_3Zr_2O_{12}}}. Values of \ce{\Delta E_{rxt}//Cathode} are reaction energies of full/half lithiated cathodes and coating interfaces. 
} 
\label{electro_window}
{\small
\begin{tabular*}{\textwidth}{@{\extracolsep{\fill}}lcccccccc@{}}
\hline \hline 
\multirow{3}{*}{\bf Coating} & \multicolumn{2}{c}{\multirow{2}{*}{{\bf Stability Window}}} & \multicolumn{4}{c}{\multirow{2}{*}{{\bf\ce{\mathbf{\Delta E_{rxt}//Cathode}}{\cite{YihanXiao2019computationalcoatingscreen}}}}} & \multicolumn{2}{c}{\multirow{2}{*}{{\bf\ce{\mathbf{\Delta E_{rxt}//SE}} {\cite{YizhouZhu2016Firstprinciplecoating}}}}}  \\ 
& \multicolumn{2}{c}{} & \multicolumn{2}{c}{} & \multicolumn{2}{c}{} \\ \cline{2-9} 
& {\bf Red.} & {\bf Ox.} & \bf{NMC} & {\bf LCO} & {\bf LMO} & {\bf LFP} & {\bf \ce{\mathbf{Li_3PS_4}}} & {\bf LLZO} \\ 
\hline
\ce{LiNbO_3} & 1.89 & 3.86 & --4/0 & 0/0 & 0/--21 & --23/--35 & --155  & --76  \\ 
\ce{Li_3NbO_4} & 1.06 & 3.58 & -- & -- &-- & -- & --132  & --4 \\ 
\ce{LiTaO_3} & 1.25 & 3.94 & 0/0 & 0/0 & 0/--7 & --20/--24 & --49  & --68 \\ 
\ce{Li_3TaO_4}  & 0.44 & 3.55 & -- & -- &-- & -- & --64   & --3 \\
\hline \hline 
\end{tabular*}
}
\end{table*}

At more negative potential ($\Delta \mu_{\mathrm{Li}}$) of Figure~{\ref{chempot_ref}}, which is more positive voltage vs.~{\ce{Li/Li^+}}, {\ce{LiNbO_3}} ({\ce{LiTaO_3}}) is in direct equilibrium with {\ce{LiNb_3O_8}} ({\ce{LiTa_3O_8}}) and in agreement with the phase diagram of Figure~{\ref{tieline_pd}}. This condition is said to be Li-poor or oxidative and can be realized, for example, when the coating material interfaces with a high-voltage cathode material, such as $>$4.20 V vs. {\ce{Li/Li^+}} in {\ce{LiNi_xMn_yCo_{1-x-y}O_2}}, NMC. The Li-poor conditions can be also mimicked by a deficiency of Li precursors during synthesis. In contrast, at more positive potentials (lower voltages vs. {\ce{Li/Li^+}}) {\ce{LiNbO_3}} ({\ce{LiTaO_3}}) is in equilibrium with {\ce{Li_3NbO_4}} ({\ce{Li_3TaO_4}}). This condition sets a chemically reducing environment (or Li-rich). From a synthesis point of view, this situation can be achieved by adding an excess of Li precursors or reacting with Li resources within cathodes. 

The computed (@\ce{r^2}SCAN) electrochemical stability windows of \ce{LiMO_3} and \ce{Li_3MO_4} are summarized in Table~\ref{electro_window}, which are in good agreement with previous reports.\cite{YizhouZhu2016Firstprinciplecoating} {\ce{LiNbO_3}}, {\ce{LiTaO_3}}, {\ce{Li_3NbO_4}}, and {\ce{Li_3TaO_4}} show relatively poor reducing (cathodic) stabilities towards low-voltage electrode materials, with the lowest voltage ($\sim$0.44~V~vs.~Li/{\ce{Li^+}}) hit by {\ce{Li_3TaO_4}}. Similarly, \ce{LiNbO_3} and \ce{LiTaO_3} show low oxidative (anodic) limits below 4~V vs.\ Li/\ce{Li^+} (see Table~\ref{electro_window}). Furthermore, the computed values of  anodic limits of \ce{Li_3NbO_4} and \ce{Li_3TaO_4} appear slightly lower than their \ce{LiMO_3} analogs. 

It is important to show whether the coating materials will mix spontaneously with the most common cathode electrodes, such as \ce{LiCoO_2} (LCO), \ce{LiMn_2O_4} (LMO), {\ce{LiNi_xMn_yCo_{1-x-y}O_2}} (NMC), and \ce{LiFePO_4} (LFP), or electrolytes, e.g., \ce{Li_3PS_4} and {\ce{Li_7La_3Zr_2O_{12}}} (LLZO). The {\ce{\Delta E_{rxt}}} of Eq.~\ref{E_rxt} is determined by the mixing ratio $x$ providing the largest enthalpy of mixing.\cite{YihanXiao2019computationalcoatingscreen}
\begin{equation}
\begin{aligned}
    \mathrm{\Delta E_{rxt}}  & = \mathrm{min}_{x\, \mathrm{in\, [0,~1]}}\{ \mathrm{E_{pd}}\left[ x\mathrm{c_a}+(1-x)\mathrm{c_b}\right] - \\   & + x\mathrm{E(c_a)} - (1-x)\mathrm{E(c_b)} \}
\end{aligned}
\label{E_rxt}
\end{equation}
where $\mathrm{E_{pd}}$, the $\mathrm{E(c_a)}$, and $\mathrm{E(c_b)}$ are the DFT energies of the mixed phase, and the electrode (or electrolyte) material $\mathrm{c_a}$, and coating material $\mathrm{c_b}$, respectively. From Table~\ref{electro_window}, niobates and tantalates exhibit very low chemical reactivity (\ce{\Delta E_{rxt}}) with most oxide-based positive   electrodes.\cite{YihanXiao2019computationalcoatingscreen} A low chemical reactivity with high-voltage electrodes is further enhanced if these niobates and tantalates are crystalline instead of amorphous.

When paired with thiophosphate SEs, i.e., \ce{Li_3PS_4} in Table~\ref{electro_window}, the large \ce{\Delta E_{rxt}} suggests that niobates may not be as stable as tantalates with thiophosphate SEs. However, the limited reactivity of tantalates compared to niobates may be explained by the lack of reported compounds containing Ta and S in contrast to more studied phases incorporating Nb and S.

\section{Intrinsic Defects in Li-Nb-O and Li-Ta-O Coatings} 
\label{DFE}
\begin{figure*}[!ht]
    \centering
    \includegraphics[width=\textwidth]{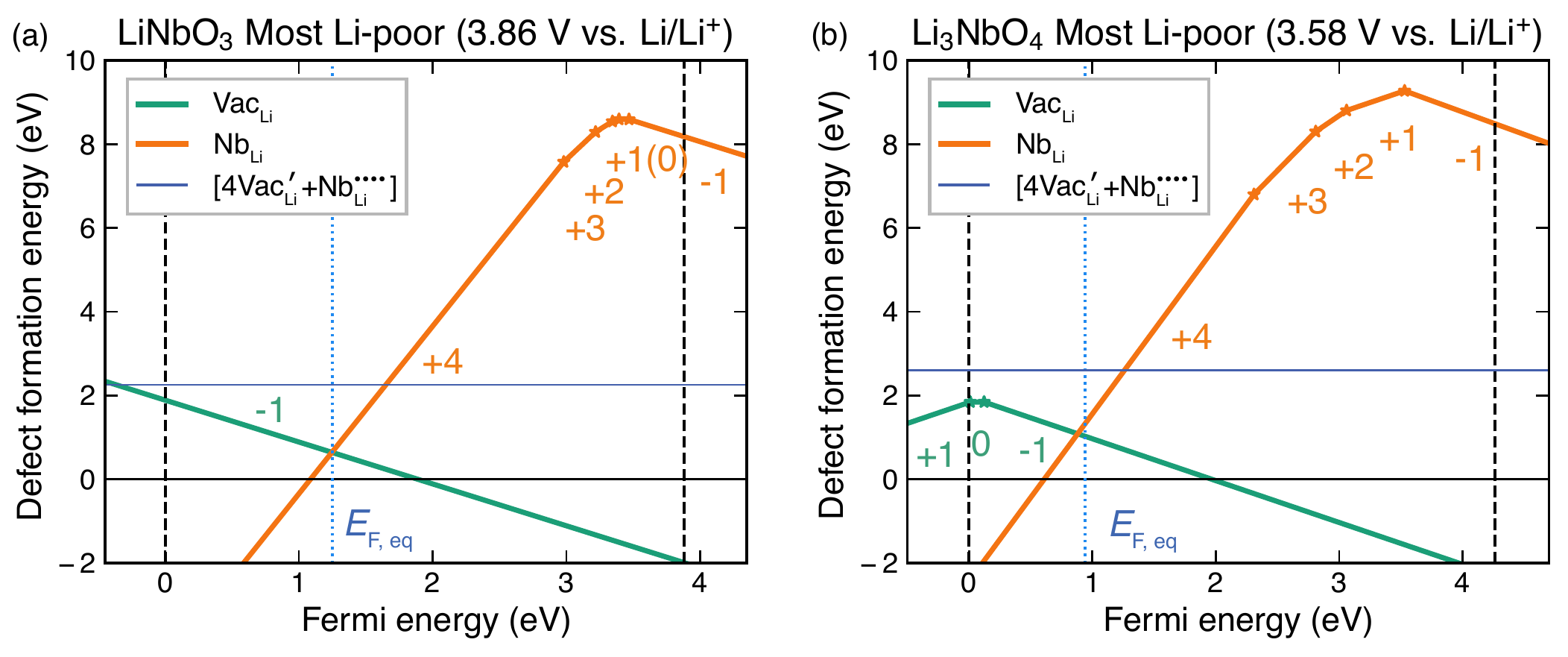}
    \caption{Computed formation energies $\Delta E_D$ of intrinsic point defects  in (a) \ce{LiNbO_3} and in (b) \ce{Li_3NbO_4} vs.\  Fermi energy, and under the most Li-poor conditions. The chemical potentials for (a) were defined by the equilibrium \ce{LiNbO_3}--\ce{LiNb_3O_8}--\ce{O_2} ($\Delta \mu_\mathrm{{Li}}=-3.86$ eV), and (b) by the \ce{Li_3NbO_4}--\ce{LiNbO_3}--\ce{O_2} equilibrium  ($\Delta \mu_\mathrm{{Li}}=-3.58$ eV). Fermi energy is referenced to the valence-band maximum (VBM). The band gaps of $\sim$3.88~eV for \ce{LiNbO_3}, and $\sim$4.26~eV for \ce{Li_3NbO_4} set maximum values of the Fermi energy.}
    \label{DFE_LNL3N_poor}
\end{figure*}

As shown in the phase diagram (Figure \ref{tieline_pd}a), the congruent composition of \ce{LiNbO_3} corresponds to $\sim$48.55 mol\% \ce{Li_2O} and 51.45 mol\% \ce{Nb_2O_5}, with a [Li]/[Nb] ratio equals to 0.94, indicating a Li deficiency of $\sim$6\% in \ce{LiNbO_3}.\cite{Scott1972expstoichiometryLNLT} Previously, the nonstoichiometry  in {\ce{LiNbO_3}} was explained by three possible defect models: (\emph{i}) lithium and oxygen vacancy, \ce{[Li]_{1-2x}Nb[O]_{3-x}}, (\emph{ii}) lithium vacancy and niobium antisite, \ce{[Li]_{1-5x}[Nb_{Li}]_xNbO_3}, (\emph{iii}) niobium vacancy and niobium antisite, \ce{Li[Nb_{Li}]_{5x}[Nb]_{1-4x}O_3}.\cite{Volk2008LNphasediagram,Smyth1983LNdefect} Model (\emph{i}) can be immediately excluded as the density of the nonstoichiometric phase was found to increase with the decreasing lithium content.\cite{Lerner1968,Iyi1992comparative} 

To date, model (\emph{ii}) is the most accepted to explain the nonstoichiometry of \ce{LiNbO_3}. X\nobreakdash-ray and neutron diffraction,\cite{Wilkinson1993defectLiNbO3,Iyi1992comparative} and solid-state nuclear magnetic resonance\cite{Blumel1994ssNMRLN} experiments of \ce{LiNbO_3} suggested that $\sim$1\% of the Li sites are occupied by Nb, and $\sim$4\% of the Li sites are vacant. This ratio of 1:4 is in line with the formation of a complex charge-neutral defect encompassing one Nb antisite defect ($\mathrm{Nb_{Li}^{\bullet\bullet\bullet\bullet}}$) accompanied by four Li vacancies ($\mathrm{{Vac}^{'}_{Li}}$), using the Kro\"ger–Vink notation. First-principles calculations confirmed that model (\emph{ii}) is preferred compared to model (\emph{iii}) under both Li-rich and Li-poor conditions.\cite{HaixuanXu2008Stability} \citeauthor{Donnerberg1989}\cite{Donnerberg1989} investigated cation vacancies in {\ce{LiNbO_3}}, and showed that the formation of Nb vacancies $\mathrm{{Vac}^{'''''}_{Nb}}$ is less favorable than Li vacancies $\mathrm{{Vac}^{'}_{Li}}$. 

As both \ce{LiNbO_3} and \ce{Li_3NbO_4} are reported in the Li--Nb--O coating layer,\cite{FengxiaXin2021WhatistheRole, Xin2021Conditioning, YantaoZhang2014} here, we extend our defect analysis to \ce{Li_3NbO_4}. Computational studies on the stability of charge-neutral defects in \ce{Li_3NbO_4} and \ce{LiNbO_3} suggested that Li/Nb antisite neutral complex and Li-Frenkel defects could exist.\cite{Kuganathan2019Li3NbO4defect, HaixuanXu2008Stability} Therefore, we investigate the stability of Li vacancies ($\mathrm{{Vac}_{Li}}$) and Nb antisites  ($\mathrm{{Nb}_{Li}}$), as well as the charge neutral complex defects $\left[4\mathrm{{Vac}^{'}_{Li}}+ \mathrm{{Nb}^{\bullet\bullet\bullet\bullet}_{Li}}\right]$ of model (\emph{ii}).

Figure~\ref{DFE_LNL3N_poor} plots the computed (@r$^2$SCAN) formation energies $\Delta E_D$ of isolated Li vacancies ($\mathrm{{Vac}_{Li}}$) and Nb antisites ($\mathrm{{Nb}_{Li}}$) in \ce{LiNbO_3} and \ce{Li_3NbO_4} following (Eq.~\ref{DFEequation}). The results of isolated point defects in {\ce{LiTaO_3}} and {\ce{Li_3TaO_4}} are provided in Figure S7, Table S8, and S9. $\Delta E_D$s are plotted as a function of Fermi energy ($E_{\mathrm{Fermi}}$). We referenced the absolute values of the Fermi energy to the energy of the valence-band maximum (VBM) of each bulk structure. The slopes of the defect lines represent different charge states of the defect ($q$ in Eq.~\ref{DFEequation}). 

High-temperature calcination facilitates the consumption of Li resources in cathodes, hence setting a condition of Li-rich for \ce{LiNbO_3} and Li-poor for \ce{Li_3NbO_4} at the coexistence of these two phases (Section~\ref{Discussion}). Upon charging, the availability of free Li in the cathode sets Li-rich conditions for the coating components (see Figure~{\ref{chempot_ref}}a). Therefore, it is crucial to investigate the formation of defects under different chemical/electrochemical conditions of the coating.

Solid lines in Figure~\ref{DFE_LNL3N_poor} represent the $\Delta E_D$s under the most Li-poor (oxidative) conditions for \ce{LiNbO_3} (\ce{LiNbO_3}--\ce{LiNb_3O_8}--\ce{O_2} equilibrium) and \ce{Li_3NbO_4} (\ce{Li_3NbO_4}--\ce{LiNbO_3}--\ce{O_2} equilibrium). The defect formation energies under the most Li-rich (reducing) conditions of \ce{LiNbO_3} and \ce{Li_3NbO_4} can be found in Figure~S6 of the SI. By definition, values of $\Delta E_D$ must be always positive quantities. In Figure~{\ref{DFE_LNL3N_poor}}a and Figure~{\ref{DFE_LNL3N_poor}}b at specific Fermi energies ($<$1~eV and $>$2.5~eV), the computed $\Delta E_D$s are negative. This indicates that these defects are unstable in the phase space set by the chemical potentials of Li and Nb (Figure~{\ref{chempot_ref}}).

Based on the temperature of calcination of coating materials ($\sim$1000K), \cite{GuozhongLu2021Study,Xin2021Conditioning,YantaoZhang2014}  both \ce{LiNbO_3} and \ce{Li_3NbO_4} phases coexist in the crystalline niobate coatings. Room temperature (300K) data are shown in Table~S6 and S7 of the SI. Vertical dotted (blue) lines are the equilibrium Fermi energy ($E_\mathrm{F,eq}$) under the Li-poor condition. Note, the equilibrium Fermi energy informs about the availability of electrons (or holes) in the coating materials.\cite{Piero2017pointdefect,LiYuheng2022pointdefect} 

Under the most Li-poor condition for \ce{LiNbO_3} (Figure~\ref{DFE_LNL3N_poor}a), the equilibrium Fermi Energy ($E_\mathrm{F,eq}$) is $\sim$1.271~eV at 1000K. The defect formation energies for $\mathrm{Vac_{Li}^{'}}$ and $\mathrm{Nb_{Li}^{\bullet\bullet\bullet\bullet}}$ are 0.62~eV and 0.74~eV, resulting in defect concentrations at $\mathrm{10^{19}~cm^{-3}}$ and $\mathrm{10^{18}~cm^{-3}}$, respectively. The carrier (hole) concentration reaches the order of $\mathrm{10^{15}~cm^{-3}}$. Under the most Li-rich condition in \ce{LiNbO_3} (Figure~S6a), $E_\mathrm{F,eq}$ is shifted to 2.89~eV at 1000K, where $\Delta E_D$ for $\mathrm{Vac_{Li}^{'}}$ and $\mathrm{Nb_{Li}^{\bullet\bullet\bullet\bullet}}$ are 0.96~eV and 1.09~eV, respectively. The corresponding defect concentrations decrease to $\mathrm{10^{17}~cm^{-3}}$ and $\mathrm{10^{16}~cm^{-3}}$. This trend shows that both Li vacancies and Nb antisites are less favored in the Li-rich condition of \ce{LiNbO_3} at the same temperature, resulting in a  decrease in defect concentrations. \ce{LiNbO_3} in Li-rich conditions displays a higher free electron concentration around $\mathrm{10^{16}~cm^{-3}}$. 

Under the most Li-poor condition in \ce{Li_3NbO_4} at 1000K (Figure~\ref{DFE_LNL3N_poor}b), $E_{F,eq}$ is at 0.93 eV. The corresponding $\Delta E_D$ for $\mathrm{Vac_{Li}^{'}}$ and $\mathrm{Nb_{Li}^{\bullet\bullet\bullet\bullet}}$ are 1.04~eV and 1.28~eV, much higher than those of \ce{LiNbO_3}. The defect concentrations of $\mathrm{Vac_{Li}^{'}}$ and $\mathrm{Nb_{Li}^{\bullet\bullet\bullet\bullet}}$ are at $\mathrm{10^{17}~cm^{-3}}$ and $\mathrm{10^{16}~cm^{-3}}$, respectively. The charge (hole) carrier concentration is $\mathrm{10^{17}~cm^{-3}}$. Under the most Li-rich condition for \ce{Li_3NbO_4} at 1000K (Figure~S6b solid lines), $E_{F,eq}$ moves to 2.85~eV, where $\Delta E_D$ for $\mathrm{Vac_{Li}^{'}}$, is 1.61~eV, and for less probable defects $\mathrm{Nb_{Li}^{\bullet\bullet}}$, and $\mathrm{Nb_{Li}^{\bullet\bullet\bullet}}$, 1.71~eV, and 1.75~eV, respectively. This indicates that \ce{Nb_{Li}} with lower charge states are likely to be more stable than $\mathrm{Nb_{Li}^{\bullet\bullet\bullet\bullet}}$ in \ce{Li_3NbO_4} under such conditions. But the defect concentrations of Li vacancies and Nb antisites are only at $\mathrm{10^{14}~cm^{-3}}$ and $\mathrm{10^{13}~cm^{-3}}$, and hence several orders of magnitude lower those in the Li-poor condition. The free electron concentration of \ce{Li_3NbO_4} in Li-rich condition is only around $\mathrm{10^{13}~cm^{-3}}$.

We now analyze the stability of neutral defect complex in {\ce{LiMO_3}} and {\ce{Li_3MO_4}} (M~=~Nb, Ta) comprising of four Li vacancies together with one Nb (Ta) antisite, thereafter indicated as $\left[4 \mathrm{Vac_{Li}^{'}}+\mathrm{M_{Li}^{\bullet\bullet\bullet\bullet}}\right]$. The spatial arrangement of antisite and Li vacancies in {\ce{LiNbO_3}} and {\ce{Li_3NbO_4}} may vary and affect the relative stability of these defects. To this end,  we investigated  a number of possibilities (spatial orderings), whose details are in Section S3 of SI. Defect complex models in \ce{Li_3TaO_4} with three unique crystallographic Li sites were constructed for each Li site. The configuration where the Li2 site is exchanged by a Ta atom shows the lowest formation energy. Table~\ref{DFEdata} summarizes the $\Delta E_D$ of lowest-energy configurations among all defect models under the Li-poor conditions.

\begin{table}[!ht]
    \caption{
    Computed r$^2$SCAN defect formation energies ($\Delta E_D$ in eV) of neutral antisite defect complexes ($\left[4 \mathrm{Vac_{Li}^{'}}+\mathrm{M_{Li}^{\bullet\bullet\bullet\bullet}}\right]$) compared to the sum of the isolated defects ($4 \mathrm{Vac_{Li}^{'}}+\mathrm{M_{Li}^{\bullet\bullet\bullet\bullet}}$, M~=~Nb, Ta). Calculations of antisite defect complexes $\left[4 \mathrm{Vac_{Li}^{'}}+\mathrm{M_{Li}^{\bullet\bullet\bullet\bullet}}\right]$ used large supercell models of \ce{LiMO_3}, (\ce{Li_{64}M_{64}O_{192}}) and \ce{Li_3MO_4} (\ce{Li_{96}M_{32}O_{128}}). The Li-poor reference states are provided in Table~S10 of the SI.
    } 
    \label{DFEdata}
    \centering
    {\small
    \begin{tabular*}{\columnwidth}{@{\extracolsep{\fill}}lcc@{}}
    \hline \hline
   {\bf Coating}  &  $\left[\mathrm{4Vac^{'}_{Li}+M^{\bullet\bullet\bullet\bullet}_{Li}}\right]$ & $\mathrm{4Vac^{'}_{Li}+M^{\bullet\bullet\bullet\bullet}_{Li}}$ \\
   \hline
    \ce{LiNbO_3}@Li-poor & 2.26 & 3.23 \\
    \ce{Li_3NbO_4}@Li-poor & 2.61 & 5.43 \\
    \ce{LiTaO_3}@Li-poor & 2.09 & 3.28 \\
    \ce{Li_3TaO_4}@Li-poor & 2.31 & 6.05 \\    
    \hline \hline
    \end{tabular*}
    }
\end{table}

As shown in Table~\ref{DFEdata}, values of $\Delta E_D$s of neutral defect complexes in {\ce{LiMO_3}} and in {\ce{Li_3MO_4}} are comparatively lower than the sum of the individual point defects (last column). For example, the sum of defect formation energies for 4 $\mathrm{Vac_{Li}^{'}}$ and $\mathrm{Nb_{Li}^{\bullet\bullet\bullet\bullet}}$ in \ce{LiNbO_3} at $E_\mathrm{{F,eq}}$ under the Li-poor condition is $\sim$3.23~eV, while that of the complex neutral defect is $\sim$2.26~eV. This indicates the favorable enthalpy gain of forming a stable complex as discussed by \citeauthor{YanluLi2015Defect}\cite{YanluLi2015Defect} In the reducing (Li-poor) conditions, the minimum $\Delta E_D$ of the defect complex in \ce{Li_3NbO_4} is $\sim$2.61~eV.

\section{Li-ion Mobility in \ce{\mathbf{LiMO_3}} and \ce{\mathbf{Li_3MO_4}} Coatings} \label{SoftBV_NEB_MTP}

In this section, we address the mobility of Li-ions in {\ce{LiMO_3}} and {\ce{Li_3MO_4}} by performing inexpensive but empirical SoftBV predictions, followed by accurate first-principles nudged elastic band (NEB) simulations.

\begin{figure}[!ht]
    \centering
    \includegraphics[width=\columnwidth]{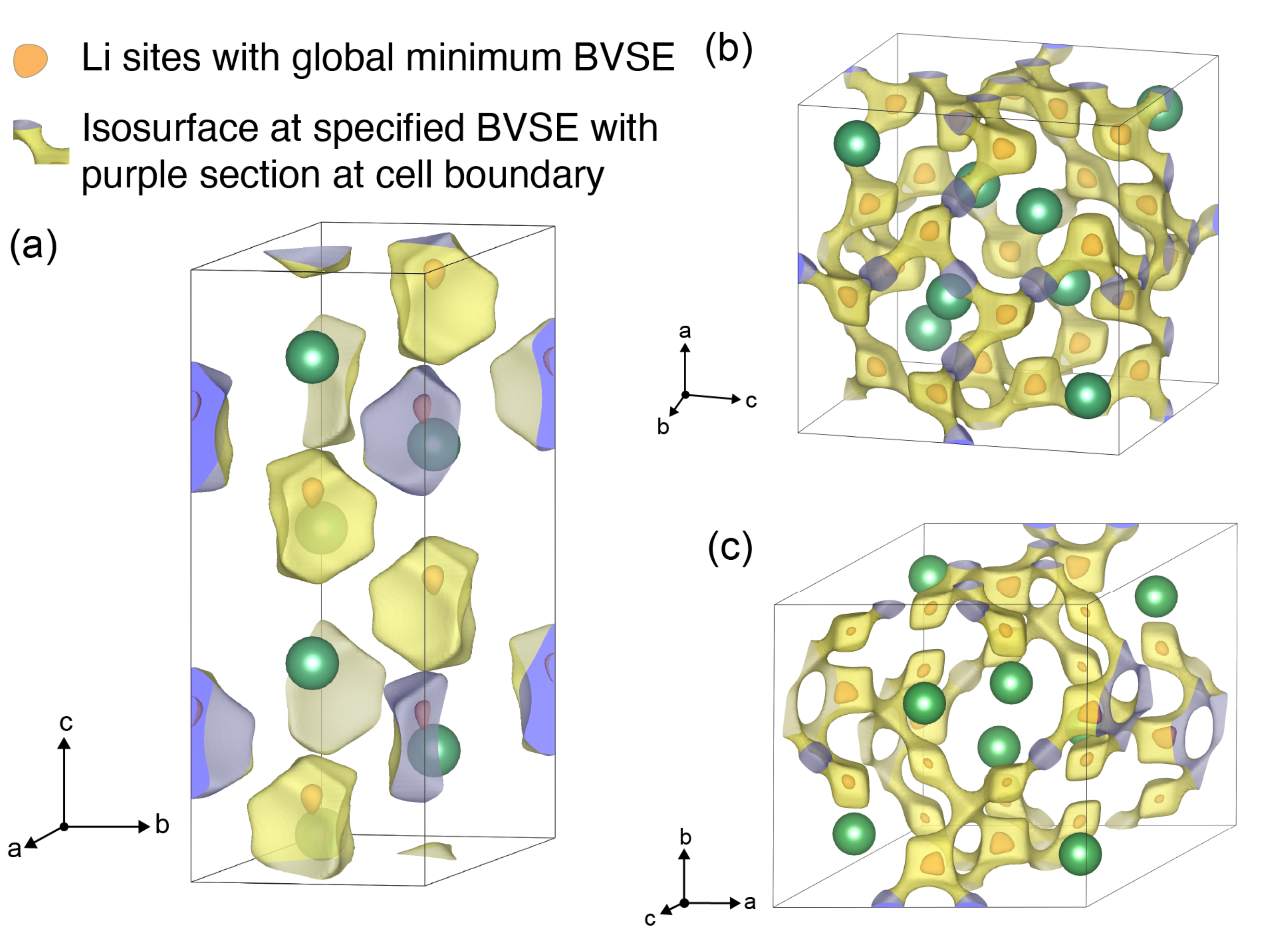}
    \caption{
    BVSE landscapes obtained from SoftBV of (a) $R3c$ \ce{LiMO_3} (M~=~Nb, Ta), (b) $I\overline{4}3m$ \ce{Li_3NbO_4}, and (c) $C2/c$ \ce{Li_3TaO_4}. Values of volumetric isosurfaces are set at 2000~meV, 700~meV, and 700~meV above the global minimum-energy set by Li sites, within \ce{LiNbO_3} (\ce{LiTaO_3}), \ce{Li_3NbO_4}, and \ce{Li_3TaO_4}, respectively. Li sites setting global minima are colored in orange. Nb and Ta atoms are shown in green and serve as a guide for the eye.
    }
    \label{softBV_BVSE}
\end{figure}

Figure~\ref{softBV_BVSE} shows the bond valence site energy (BVSE) landscape computed with SoftBV for each bulk structure, whose volumetric isosurfaces are set to 2000~meV (in \ce{LiMO_3}) and 700~meV (in\ce{Li_3MO_4}), above the global minimum of Li-site energy computed by SoftBV. From the disconnected isosurfaces seen in Figure~\ref{softBV_BVSE}a, it is unlikely for Li-ions to migrate in the defect-free crystalline phases of \ce{LiNbO_3} or \ce{LiTaO_3}, as no Li``channels" are formed between adjacent sites. Figure~\ref{softBV_BVSE}b and Figure~S8 in the SI clearly depict extended percolating Li-diffusion channels of \ce{Li_3NbO_4}. For the ordered $\beta$-phase \ce{Li_3TaO_4}, Figure~\ref{softBV_BVSE}c and Figure~S9 show the percolating networks of Li ions. Insightful illustrations of the migration environments near the crystallographic Li1, Li2, and Li3 sites are in Figure~S13a, b, and c.

From the SoftBV analysis, the percolating migration barriers --- that give rise to a net transport of Li-ions within the coating material --- were further investigated using NEBs coupled with DFT@r$^2$SCAN. Three main types of {\ce{Li^+}}-migration pathway can be  envisioned:
\begin{enumerate}
    \item {\bf Pathway}~\ce{\mathbf{Vac_{Li}}}: One Li vacancy $\mathrm{Vac_{Li}^{'}}$ (low-vacancy limit) was introduced  and compensated by a background charge (in the form of jellium). 
    \item {\bf Pathway}~\ce{\mathbf{4Vac_{Li}+M_{Li}}} (M~=~Nb, Ta): One Li vacancy $\mathrm{Vac_{Li}^{'}}$ of the four vacancies of the charge-neutral defect complex $\left[\mathrm{4Vac^{'}_{Li}+M^{\bullet\bullet\bullet\bullet}_{Li}}\right]$ migrates. 
    \item {\bf Pathway}~\ce{\mathbf{4Vac^\triangle_{Li}+M_{Li}}}: One Li vacancy $\mathrm{Vac_{Li}^{'}}$ of the four vacancies of the charge-neutral complex defect $\left[\mathrm{4Vac^{'}_{Li}+M^{\bullet\bullet\bullet\bullet}_{Li}}\right]$ migrates. All the other three $\mathrm{Vac_{Li}^{'}}$ were located near the migration event.
\end{enumerate}

\begin{figure*}[!ht]
    \centering
    \includegraphics[width=\textwidth]{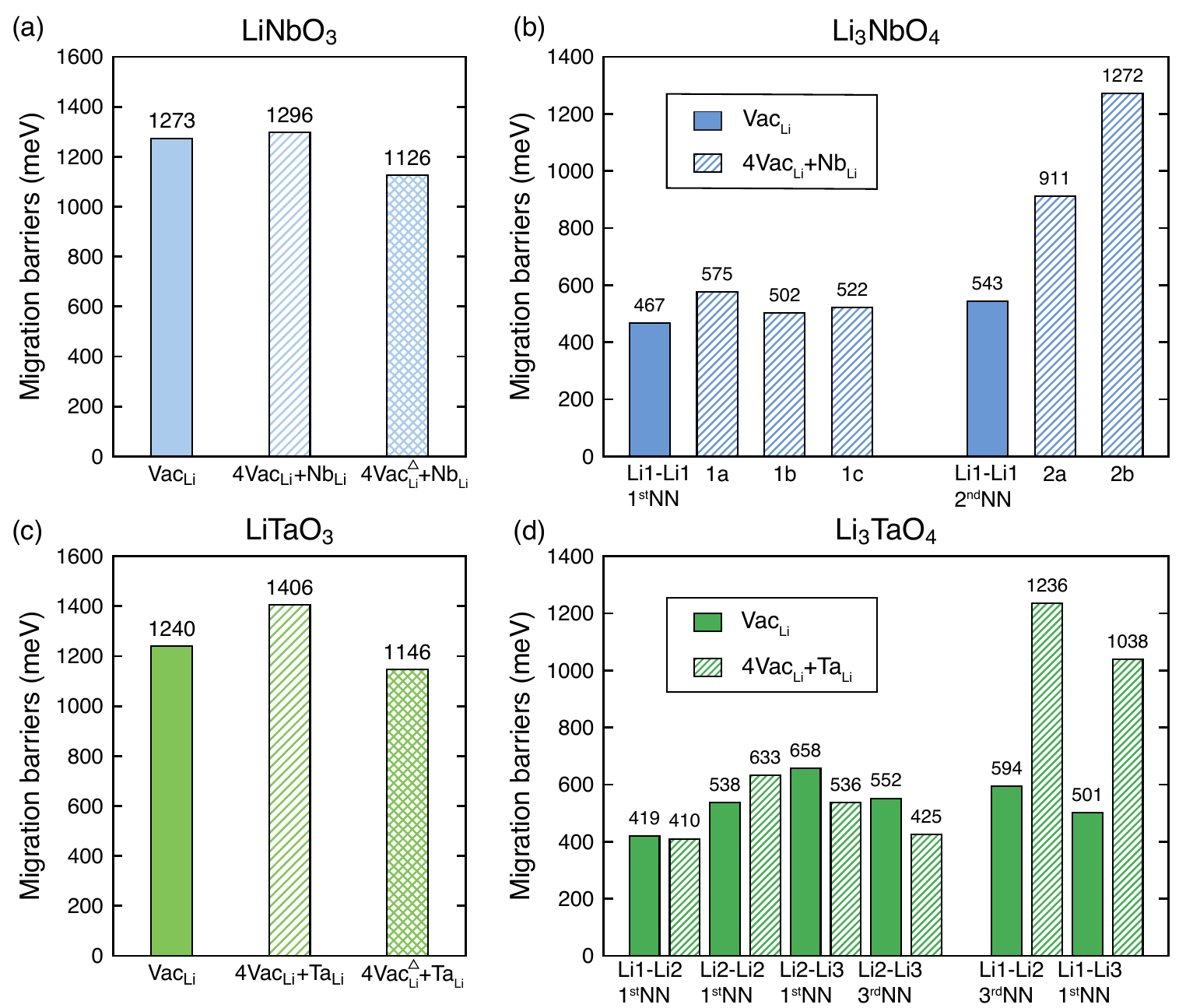}
    \caption{
    Li-ion migration barriers computed with NEB@r$^2$SCAN in (a) \ce{LiNbO_3}, (b) \ce{Li_3NbO_4}, (c) \ce{LiTaO_3}, and (d) \ce{Li_3TaO_4} supercell structures. In these structures, the environment of Li-ions is set in terms of near neighbors, NNs (Figure~{\ref{structure_LNL3N}} and Figure~{\ref{structure_L3T}}). Filled bars show migration barriers computed in the low-vacancy limit (incorporating one $\mathrm{Vac^{'}_{Li}}$ per supercell).  Hatched bars show migration barriers of models incorporating neutral complex antisite-vacancy defects ($\mathrm{4Vac_{Li}+M_{Li}}$). Bars with crossed patterns show migration barriers for {\ce{LiMO_3}} in the presence of antisite defect complex, with all Li vacancies near the migration pathways and indicated as \ce{4Vac^\triangle_{Li}+M_{Li}}. Configurations of all pathways are provided in Figure~S11-S13.
    \label{NEB}
    }
\end{figure*}

Based on these three distinct pathways, distinct migration models were identified by looking at the coordination environment of the migrating Li-ion (Figure~{\ref{structure_LNL3N}} and Figure~{\ref{structure_L3T}}). Values of computed migration barriers are summarized in Figure~\ref{NEB} and in Table~S11 of the SI. 

From Figure~{\ref{NEB}}a and Figure~{\ref{NEB}}c, no matter which of the three pathways are chosen, the predicted migration barriers in \ce{LiNbO_3} and \ce{LiTaO_3} are consistently larger than 1100~meV. This data clearly demonstrates that Li-ion transport in {\ce{LiNbO_3}} and {\ce{LiTaO_3}} relying on intrinsic defects cannot guarantee sufficient Li-ion transport.

In particular, for {\ce{LiNbO_3}} and {\ce{LiTaO_3}}, the Li-ion migration barriers in the low-vacancy limit model ({\ce{Vac_{Li}}}) are 1273~meV and 1240~meV, respectively, in agreement with previous experimental results.{\cite{Glass1978Ionic,Heitjans2006NMRsingle}} To investigate whether the presence of neutral defect complex ($\left[\mathrm{4Vac^{'}_{Li}+M^{\bullet\bullet\bullet\bullet}_{Li}}\right]$) can lower the high migration barriers of crystalline \ce{LiNbO_3} (\ce{LiTaO_3}), we studied the Li-ion migration in the vicinity of these defects (see Figure~S11b and S11c). As shown in Figure~\ref{NEB}a and Figure~\ref{NEB}c, migration barriers are still as high as $\sim$1296~meV for \ce{LiNbO_3} and $\sim$1406~meV for \ce{LiTaO_3}. The computed migration energies of Li-ion in congruent \ce{LiNbO_3} ($\sim$1296~meV), appear in excellent agreement with experimental values ($\sim$1300~meV).\cite{Rahn2011congruentLNconductivity, Huger2014singleLiNbO3Ea}   

Starting from the hypothesis that the presence of Li vacancies near the migrating Li-ion may lower its migration barriers, in Li migration models labeled as $\mathrm{4Vac_{Li}^\triangle+Ta_{Li}}$, we forced all four Li vacancies to be close to the Li migration pathway (Figure~S11d). Although the activation barriers ($\sim$1126~meV for \ce{LiNbO_3} and $\sim$1146~meV in \ce{LiTaO_3}) are slightly lower than in the other cases, such barriers are still too high to achieve reasonable Li-ion transport across the coating layer. 

\ce{Li_3NbO_4} and \ce{Li_3TaO_4} exhibit much more promising transport properties with lower activation barriers, on average lower than 700~meV (Figure~\ref{NEB}b and Figure~\ref{NEB}d). For example, in \ce{Li_3NbO_4}, one Li vacancy can migrate to its two \nth{1}NNs ($\sim$467 meV), four \nth{2}NNs ($\sim$543 meV), and one \nth{3}NN ($\sim$701 meV, shown in Figure~S12b). Among 11 distinct pathways considered between three types of Li sites in \ce{Li_3TaO_4}, eight migration barriers are consistently lower than 700~meV (see Table~S11). Therefore, in {\ce{Li_3NbO_4}} and {\ce{Li_3TaO_4}}, several migration pathways may enable macroscopic diffusion of Li-ions (Figure~S8 and S9).

In analogy to {\ce{LiMO_3}}, we also explore the Li migration in {\ce{Li_3MO_4}} (M~=~Nb, Ta) in the vicinity of the {\ce{M_{Li}}} antisite and four Li vacancies. The computed paths of these Li migrations are shown in Figure~S12d, Figure~S13d--f, respectively. From Figure \ref{NEB}b, in \ce{Li_3NbO_4}, the \ce{Li}-ion migration barriers generally increase in the proximity of Nb antisite  defects  when compared to the migration of a diluted Li vacancy (models {\ce{Vac_{Li}}} in Figure~{\ref{NEB}}b). However, most pathways in Figure~{\ref{NEB}}b still show relatively low  barriers (model 1a, model 1b, and model 1c with barriers of 535~meV, 502~meV, and 522~meV). A similar trend is found for \ce{Li_3TaO_4} as depicted in Figure~{\ref{NEB}}c. In both {\ce{Li_3NbO_4}} and {\ce{Li_3TaO_4}} specific arrangement of Li vacancies can lead to a sudden increase in the migration barriers, even beyond 1200~meV (see Figure~{\ref{NEB}}b and Figure~{\ref{NEB}}c).

\begin{figure*}[!ht]
    \centering
    \includegraphics[width=\textwidth ]{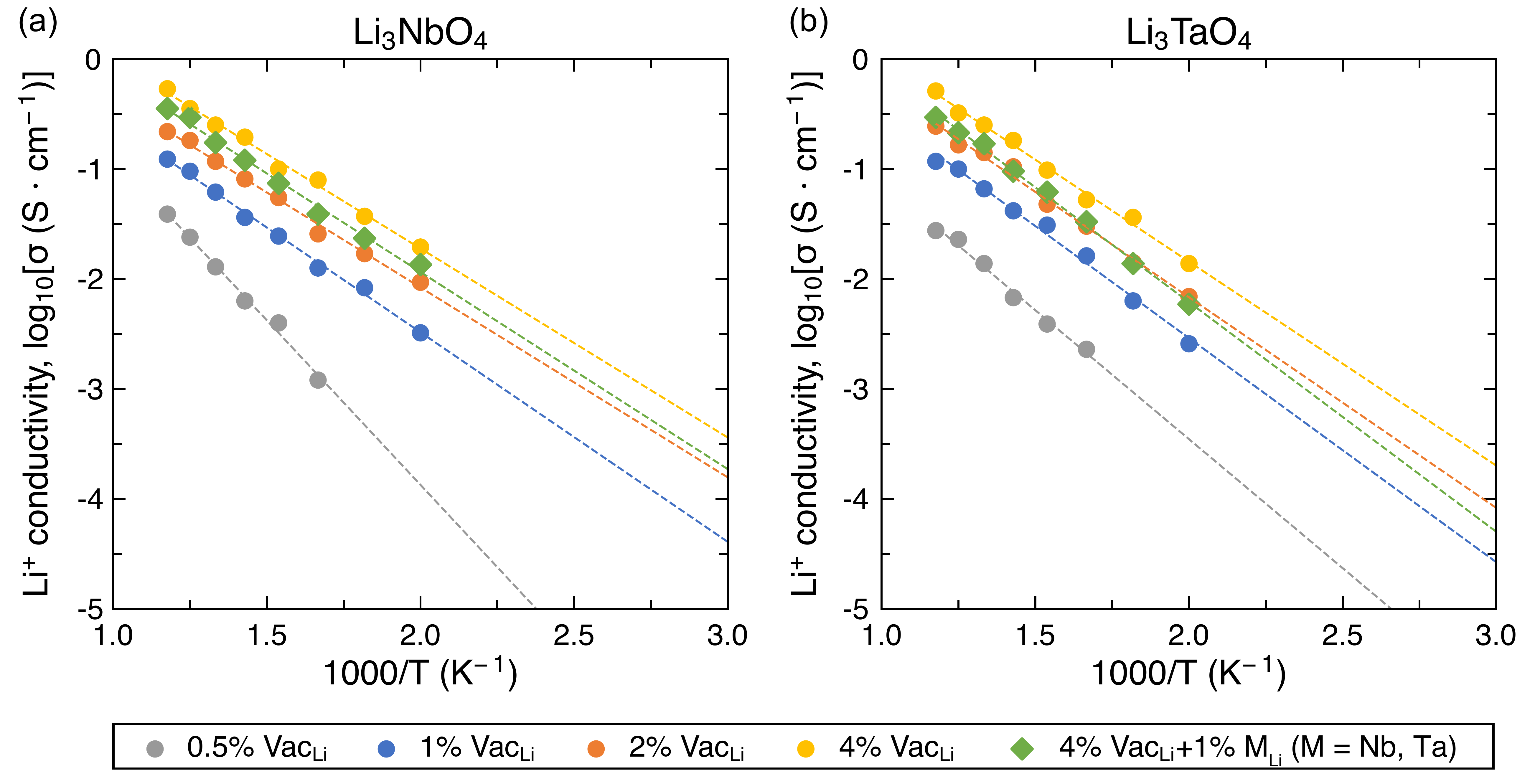}
    \caption{Arrhenius plots of Li-ion conductivities in (a) \ce{Li_3NbO_4} and (b) \ce{Li_3TaO_4} obtained from MTP-MDs trained on different defect types and concentrations. The temperatures modeled range from 500K to 850K, except for model 0.5\%. Dashed lines are the Arrhenius fits, whose migration energies are in Table~S13 of SI.
    }
    \label{MTP-MD_results}
\end{figure*}

This variation in the Li migration barriers is introduced by the immediate crystallographic environment of niobates (tantalates) and the proximity of the Li migration event to the complex antisite defect. In {\ce{LiMO_3}} (M~=~Nb, Ta), the NEB barriers for Li ions migrating in the presence of a complex defect vary slightly. This is because the closely packed structures of {\ce{LiMO_3}} are very symmetric and structural changes introduced by point defects cause minimal variations of the migration environment,  and hence the Li-ion migrating barriers. In {\ce{Li_3MO_4}}, the migration barriers of a Li vacancy (low-vacancy limit model) appear very sensitive to the migration environment as reflected in Figure~{\ref{NEB}}b and d.  In the structures of {\ce{Li_3NbO_4}} (tetramer of Nb octahedra) and {\ce{Li_3TaO_4}} (zigzag chain of Ta octahedra) of Figure {\ref{structure_LNL3N}}b and {\ref{structure_L3T}}, the high-valence Nb$^{5+}$, and Ta$^{5+}$ ions arrange sparsely, in a way to provide additional voids enabling facile {\ce{Li^+}} migration. In addition to the local environment, the introduction of antisite defect complex will affect the Li migration pathways by either reducing the {\ce{Li^+}}--{\ce{Li^+}} repulsion near the 4 Li vacancies or increasing the repulsion for pairs {\ce{Li^+}}--{\ce{M^{5+}_{Li}}}. 

So far, the computed migration barriers of Figure~{\ref{NEB}}, \ce{Li_3MO_4} (M~=~Nb, Ta) suggest facile Li-ion transport compared to \ce{LiMO_3}. To gain more insight into the complex behavior of Li-ion transport in \ce{Li_3MO_4} (M~=~Nb, Ta), we perform machine-learning molecular dynamics (MDs) simulations based on the moment tensor potential (MTP) method.\cite{MTP2019KonstantinGubaev} In the MTP, we explore the effect of different concentrations of {\ce{Vac^{'}_{Li}}}, set to 0.5\%, 1\%, 2\%, 4\%, respectively. As calculated in Section~\ref{DFE}, the $\Delta E_D$s of neutral defect complexes in \ce{Li_3MO_4} are comparable with \ce{LiMO_3}. Here, we only considered the case of 4\% Li vacancies and 1\% M antisites in \ce{Li_3MO_4}, with the MTP-MD simulations performed in the most favorable  defect models of {\ce{Li_3NbO_4}} and {\ce{Li_3TaO_4}} (see Figure S12c and S13e in SI).

Figure~\ref{MTP-MD_results} shows the Li-ion transport properties from MTP-MD simulations. Predicted migration barriers can be derived from the Arrhenius fits of Figure~\ref{MTP-MD_results}, shown in Table S13 of SI. In Figure~\ref{MTP-MD_results}, by including 0.5\% Li vacancies in \ce{Li_3NbO_4}, we achieved migration barriers $\sim$622~meV, in agreement with experimental values ($\sim$580--850~meV) of crystalline \ce{Li_3NbO_4} and NEB barriers of Figure~{\ref{NEB}}.\cite{2010Li3NbO4NMRultraslow,Glass1978Ionic,Liao2013L3NO4barrierWsubstitution} Unsurprisingly, as the concentration of Li vacancies in \ce{Li_3NbO_4} increases, the inferred migration barriers of \ce{Li_3NbO_4} of 2\% and 4\% Li vacancies decrease significantly to $\sim$400 meV. Likewise, the Li-ion conductivities increase greatly by three to five orders of magnitude by increasing the {\ce{Vac_{Li}}} from 0.5\% to 4.0\%. As shown in Figure~\ref{MTP-MD_results}b, the ionic conductivities of \ce{Li_3TaO_4} are similar to those of \ce{Li_3NbO_4}. The extrapolated Li-migration barrier in \ce{Li_3TaO_4} with 0.5\% Li vacancies is estimated to $\sim$527~meV. in agreement with the experimental measurements ($\sim$550~meV in \ce{Li_{2.98}Ta_{1.004}O_4}).\cite{ChaeeunK2023L3TaO4defectconductivity} Similar with \ce{Li_3NbO_4}, as the concentration of Li vacancies increases, \ce{Li_3TaO_4} exhibits improved conduction properties.
 
In the presence of Nb or Ta antisite defect complexes, the Li-ion conduction of \ce{Li_3MO_4} of 4\% Li vacancies and 1\% M antisites is comparable with the ionic conductivity predicted in a model system with 4\% Li vacancies. These results show that the {\ce{M_{Li}}} antisite defects may not significantly affect Li-ion transport in {\ce{Li_3MO_4}}, but {\ce{M_{Li}}} defects ensure the availability of Li vacancies. The migration barriers of Figure~{\ref{NEB}} support these observations.

\section{Discussion}
\label{Discussion}
To improve the cathode-electrolyte interfacial stability, extend the oxidation potential of electrolytes, and increase the longevity of lithium-ion batteries, coating materials are usually applied to positive electrode materials.\cite{Wang2015interfacestabilityLIBs} Commonly used coating materials are niobates and tantalates.\cite{Culver2019Funcionalitycoating} 

Niobate coating materials are typically synthesized using wet-chemical protocols. In these syntheses (see Table~S1 in the SI), the temperature of calcination affects mainly the crystallinity of {\ce{LiNbO_3}}, and the distribution of {\ce{LiNbO_3}} with secondary phases, as revealed by X-ray diffraction, and transmission electron microscopy experiments. \cite{YantaoZhang2014, FengxiaXin2021WhatistheRole, GuozhongLu2021Study}

As temperature increases, up to four ``stages'' of calcination may be involved in the fabrication of these Li--Nb--O materials (Table~S1 in the SI),\cite{NarumiOhta2007,Xin2021Conditioning, GuozhongLu2021Study,Hsiao2010L3Nbandgap} which are: 
\begin{enumerate}
        \item At low calcination temperatures ($<$400~\celsius), there only exists amorphous \ce{LiNbO_3} phase, which display high Li-ion conductivities ($10^{-5}$--10$^{-6}$~S~ cm$^{-1}$).
        \item At intermediate temperatures (400--650~\celsius), crystalline \ce{LiNbO_3} phase co-exists with the disorderd rock-salt \ce{Li_3NbO_4} phase.
        \item At high temperatures (650--750~\celsius), a mixture of crystalline \ce{LiNbO_3} phase with the ordered $I\overline{4}3m$ \ce{Li_3NbO_4} phase is identified.
        \item At temperature exceeding 750~\celsius, the niobate coating is composed of a single ordered $I\overline{4}3m$ \ce{Li_3NbO_4} phase.
        Therefore, at this stage, all \ce{LiNbO_3} phase is transformed into \ce{Li_3NbO_4}.
\end{enumerate}
The conversion temperatures between the Li--Nb--O phases mentioned above can have slight variations when the amount of precursors used in the synthesis and the calcination time vary. As an example, in the recent work of \citeauthor{Xin2021Conditioning}\cite{Xin2021Conditioning} the niobate coating materials contains only \ce{Li_3NbO_4}  when calcinated for three hours at $\sim$690~\celsius. In contrast, \citeauthor{YantaoZhang2014}\cite{YantaoZhang2014} characterized that a Li--Nb--O coating still displays the crystalline \ce{LiNbO_3}  and \ce{Li_3NbO_4} phases after calcination at $\sim$700~\celsius~for ten hours. This analysis suggests that niobate coating materials, commonly accepted as ``{\ce{LiNbO_3}}'' may co-exist with Li-rich materials, such as  {\ce{Li_3NbO_4}} (see Figure~\ref{tieline_pd}a and Figure~\ref{chempot_ref}a). From our simulations, we extend these findings to tantalate coatings that may contain varying mixtures of \ce{LiTaO_3} and \ce{Li_3TaO_4}. 

From this investigation emerged that selected calcination temperatures control the coexistence of {\ce{LiMO_3}} and {\ce{Li_3MO_4}} ({\ce{M}}~=~{\ce{Nb}}, and {\ce{Ta}}) phases in the coating materials. The concurrence of {\ce{LiMO_3}} and {\ce{Li_3MO_4}} phases can promote the formation of defects within {\ce{Li_3MO_4}} with promising ionic conductivities. Concomitantly, the presence of {\ce{LiMO_3}} extends the oxidative resistance to higher voltages, improving batteries{\textquotesingle} electrochemical performance. Nevertheless, if calcination temperatures exceed 750~{\degree C}, {\ce{LiMO_3}} will be fully depleted, and the exceeding {\ce{Li_3MO_4}} will lower the onset voltage for oxidation to $\sim$3.6V~vs~Li/{\ce{Li^+}}.

The coexistence of {\ce{LiMO_3}} and {\ce{Li_3MO_4}} suggests that the functional properties of these coatings materials, such as, (\emph{i}) the \ce{Li}-ion transport in these phases, (\emph{ii}) the chemical and electrochemical stabilities (under an external bias), and (\emph{iii}) the mechanical properties (not investigated here) are broadly linked to the multiphased nature of these materials. We built robust thermodynamic and kinetic models using state-of-the-art first-principles calculations to link the multiphased nature of niobates (tantalates) with their functional properties. \break

\noindent {\bf Intrinsic defects in coating materials:} Using DFT (at the r$^2$SCAN level of theory), we verified the natural occurrence of intrinsic defects in these materials. The equilibrium of {\ce{LiNbO_3}} and {\ce{Li_3NbO_4}} in niobates will set a Li-poor (more oxidative) condition, highly favoring the formation of neutral antisite defect complex in \ce{Li_3NbO_4}, implying the simultaneous formation of four Li vacancies and a charged \ce{Nb^{\bullet \bullet \bullet \bullet}_{Li}} antisite defect, i.e., $\left[\mathrm{4Vac^{'}_{Li}+Nb^{\bullet\bullet\bullet\bullet}_{Li}}\right]$ (Table~\ref{DFEdata} and Table~S10). We arrived at similar findings in \ce{Li_3TaO_4}, suggesting that the coexistence of \ce{LiTaO_3} and \ce{Li_3TaO_4} (Li-poor condition) will favor the formation of neutral defect complex in \ce{Li_3TaO_4}, i.e., $\left[\mathrm{4Vac^{'}_{Li}+Ta^{\bullet\bullet\bullet\bullet}_{Li}}\right]$.   

In general, in {\ce{LiNbO_3}} our DFT defect formation energies suggest that neutral defect complex, i.e., $\left[\mathrm{4Vac^{'}_{Li}+Nb^{\bullet\bullet\bullet\bullet}_{Li}}\right]$ are highly favored in the Li-poor condition. Isolated Li vacancies and \ce{Nb^{\bullet\bullet\bullet\bullet}_{Li}} antisites are also likely to form in  {\ce{LiNbO_3}}, with concentrations  at 1000K $\sim$1.39$\times10^{19}$ cm$^{-3}$, and $\sim$3.46$\times10^{18}$ cm$^{-3}$, respectively. Characterization with Neutron and X-ray diffraction,\cite{Wilkinson1993defectLiNbO3,Iyi1992comparative} and later solid-state nuclear magnetic resonance\cite{Blumel1994ssNMRLN} on {\ce{LiNbO_3}} suggested that $\sim$1\% of the Li sites are occupied by Nb, and $\sim$4\% of the Li sites are vacant, hence supporting the existence of neutral antisite defect complex and in excellent agreement with our computational models. 
While at the cathode interface, exchange of Nb (Ta) from the coating layer with transition metals of the positive electrode is possible,\cite{XinFengxia2022Nbmechanical} here, we did not investigate the occurrence of these antisite defects. \break

\noindent {\bf Li-ion transport in niobate and tantalate coatings:} 
Subsequently, we inspected the properties of Li-ion transport (with SoftBV, NEB, and MTP-MD simulations) in the presence of the defects types favored by the thermodynamic conditions in \ce{LiNbO_3}  and \ce{Li_3NbO_4}, and \ce{LiTaO_3}  and \ce{Li_3TaO_4}. In this work, we did not focus on simulating the behavior of amorphous niobates and tantalates. However, experimental investigations by \citeauthor{Rahn2011congruentLNconductivity}\cite{Rahn2011congruentLNconductivity} suggested that amorphous {\ce{LiNbO_3}} synthesized in Li-poor (more oxidative) conditions exhibits better Li-ion transport ({\ce{E_a}}~=~700~meV) than amorphous {\ce{LiNbO_3}} ({\ce{E_a}}~=~830~meV) synthesized in Li-rich conditions. This observation matches our computational results, with {\ce{LiNbO_3}} and {\ce{Li_3NbO_4}} likely to be highly defective in Li-poor conditions.

Clearly, coating materials are effective at migrating Li-ion if the migration barriers are within some prescribed tolerance values. Previously, the maximum tolerable migration barriers ($\mathrm{E^{max}_a}$) for Li-ions in coating materials was estimated using the model proposed by \citeauthor{TinaChen2019coatingthickness}\cite{TinaChen2019coatingthickness}, suggesting that coating layers of different thicknesses are only viable when the Li-migration barriers are smaller than a maximum value $\mathrm{E^{max}_a}$. Based on manufacturing data of crystalline Li--Nb--O and Li--Ta--O coatings (Table~S1), we identified ``reasonable'' operating conditions to derive a value of $\mathrm{E^{max}_a}$, which set a 10nm coating thickness, operating at 25~\celsius{} and at a discharge rate of 1C (1 hour). This yields to an $\mathrm{E^{max}_a}$ of $\sim$700~meV.

In crystalline \ce{LiNbO_3} and \ce{LiTaO_3}, our data suggest that the migration of Li-vacancies (in the dilute regimes) remain high ($>$1100~meV) with or without the presence of $\left[\mathrm{4Vac^{'}_{Li}+Nb^{\bullet\bullet\bullet\bullet}_{Li}}\right]$ in the materials. These values feature consistently above the maximum tolerable barriers, $\sim$700~meV for these coating materials. However, experimentally amorphous \ce{LiNbO_3} and \ce{LiTaO_3} both appear decent Li-ion conductors at room temperature ($\sim$10$^{-5}$--10$^{-6}$~S~cm$^{-1}$). Therefore, future computational studies are required to elucidate the Li-ion transport properties of these amorphous phases. 

The computed migration barriers in crystalline \ce{Li_3NbO_4} (or \ce{Li_3TaO_4}) phase are significantly lower (mostly $<$700~meV) than crystalline \ce{LiNbO_3} (or \ce{LiTaO_3}). Li-ion conductivities obtained with machine-learned MDs also corroborate this finding. We claim that \ce{Li_3NbO_4} or \ce{Li_3TaO_4} are the active components in niobate or tantalate coating materials. Therefore, synthesis protocols and defect engineering remain crucial to the overall properties of these coating materials. \break 

\noindent {\bf Low oxidative stability limits of \ce{\mathbf{ LiMO_3}} and \ce{\mathbf{Li_3MO_4}}:} The high operating  voltages $\geq 4.2$ V vs.\ Li/\ce{Li^+} deliver by high-Nickel content layered materials, such as the \ce{LiNi_{0.8}Co_{0.1}Mn_{0.1}O_2} (NMC811), require coating materials with matched oxidative stability windows. In Table~\ref{electro_window} we estimated the oxidative (anodic) stabilities of all phases in niobate and tantalate coating materials. 
As widely applied coating materials, \ce{LiNbO_3} and \ce{LiTaO_3} show low oxidative (anodic) limits below 4~V vs.\ Li/\ce{Li^+} (see Table~\ref{electro_window}), which are lower than reported values of some phosphate coatings, such as \ce{Li_3PO_4} ($\sim$4.2 V vs.\ Li/\ce{Li^+}) and \ce{LiPO_3} ($\sim$5.0 V).\cite{YizhouZhu2016Firstprinciplecoating} Predicted anodic limits of \ce{Li_3NbO_4} and \ce{Li_3TaO_4} appear slightly lower than their \ce{LiMO_3} analogs. Indeed, the higher Li-ion transport in crystalline \ce{Li_3NbO_4} (and \ce{Li_3TaO_4}) phases compared to \ce{LiNbO_3} (and \ce{LiTaO_3}) can explain the limiting oxidative st abilities of the \ce{Li_3MO_4} materials. Therefore, the selection of adequate calcination treatments for niobate and tantalate coating should always ensure the concurrence of the highly conductive \ce{Li_3MO_4} component and \ce{LiMO_3} phases with better anodic stabilities. 

Furthermore, from our calculations of intrinsic defects in the niobate and tantalate phases, the equilibrium Fermi energies,  {\ce{E_{F,eq}}}, (see Figure~{\ref{DFE_LNL3N_poor}}) can be linked to the electronic conductivity of these bulk materials.{\cite{Gorai2021Defect,LiYuheng2022pointdefect}} Across all the chemical and electrochemical environments (reducing or oxidizing) explored here, the equilibrium Fermi energies of lithium niobates and tantalates are always at least $\sim$1~eV away from the valence or conduction bands. This indicates that intrinsic defects in lithium niobates and tantalates bulks alone may not impart significant electronic conductivities.

\section{Conclusions}
We investigated the structure-property relationship of lithium niobates and tantalates, which are an important class of coating materials for lithium-ion batteries. We underscored the importance of the muliphasic nature of niobate coating materials, \ce{LiNbO_3} and \ce{Li_3NbO_4}. Similarly, tantalates appear in mixtures of \ce{LiTaO_3} and \ce{Li_3TaO_4}. 

We demonstrate that \ce{LiNbO_3} and \ce{Li_3NbO_4} may contain high concentrations of lithium vacancies \ce{Vac^{'}_{Li}} and antisites \ce{Nb^{\bullet \bullet \bullet \bullet}_{Li}} arranged into charge-neutral defect complexes,  $\left[\mathrm{4Vac^{'}_{Li}+M^{\bullet\bullet\bullet\bullet}_{Li}}\right]$. Our simulations reveal poor Li-ion transport of crystalline \ce{LiNbO_3} and \ce{LiTaO_3} even in the presence of intrinsic defects \ce{Vac^{'}_{Li}} and neutral antisite defect complex  $\left[\mathrm{4Vac^{'}_{Li}+M^{\bullet\bullet\bullet\bullet}_{Li}}\right]$. In contrast, our analysis demonstrates that the secondary phases \ce{Li_3NbO_4} and \ce{Li_3TaO_4} are much better ion conductors than \ce{LiNbO_3} and \ce{LiTaO_3}. 

Finally, the low-oxidative (anodic) voltages of \ce{Li_3NbO_4} and \ce{Li_3TaO_4} compared to \ce{LiNbO_3} and \ce{LiTaO_3} encourage the selection of adequate calcination treatments to ensure the concurrence of both phases. 

Our findings stimulate more in-depth investigations on the structure-property relationships of these highly sought coating materials. 


\section{Methods}
\subsection{First-principles calculations} \label{DFT}
Density functional theory (DFT) calculations were performed using the Vienna \emph{ab initio} Simulation Package.\cite{Kresse1996VASPCondens} The projector augmented wave potentials  described the core electrons, and are: Li 17Jan2003 $\mathrm{2s^1}$, Nb{\_}pv 08Apr2002 $\mathrm{4p^6 5s^1 4d^4}$, Ta{\_}pv 07Sep2000 $\mathrm{5p^6 6s^2 5d^3}$, and O 08Apr2002 $\mathrm{2s^2 2p^4}$. \cite{Kresse1999PAWpotential} The kinetic energy cut-off describing the valence electrons was set to 520~eV. The meta-GGA functional \ce{r^2}SCAN approximates the exchange-correlation energy.\cite{Furness2020R2scan} The PBE functional was used for training the machine-learned potentials,\cite{Perdew1996PBE}  as it is approximately 10\% faster than {\ce{r^2}}SCAN in the AIMDs. 5$\times$5$\times$5 \ce{\Gamma}-centered \textit{k}-point meshes were used to integrate the \nth{1} Brillouin zone of \ce{LiNbO_3} and \ce{LiTaO_3} (with a $R3c$ space group), and converged within $10^{-5}$~eV. $k$-point meshes of 3$\times$3$\times$3 and 4$\times$4$\times$2 were applied to \ce{Li_3NbO_4} ($I\overline{4}3m$) and \ce{Li_3TaO_4} ($C2/c$). In larger supercells and metals (Li, Nb, and Ta) the $k$-point meshes were optimized. With the exception of defect calculations, volumes, shapes, and coordinates of each structure were optimized until the inter-atomic forces were $\leq10^{-2}$ eV/{\AA}. 

All crystal structures were taken from the Inorganic Chemical Structure Database. Disordered rock-salt structures of {\ce{Li_3NbO_4}} and {\ce{Li_3TaO_4}} were ordered by computing all possible Li/Nb (Li/Ta) arrangements at given supercell sizes.\cite{ShyuePingOng2013pymatgen} \ce{O_2} gas was used a reference for the oxygen $\mu_O$, and was simulated by placing an oxygen dimer in a {10$\times$10$\times$10~\AA$^{3}$} box. We included {\ce{O_2}} vibrational contributions ($\sim$100~meV at 0~K).

As shown in Table~S2 and S3 of the SI, the DFT-calculated structures of {\ce{LiNbO_3}} ($R3c$), {\ce{LiTaO_3}} ($R3c$), {\ce{Li_3NbO_4}} ($I{\overline{4}}3m$), and $\beta$-{\ce{Li_3TaO4}} ($C2/c$), using the meta-GGA r$^2$SCAN exchange and correlation functional are in much better agreement with existing experimental data on the lattice parameters and electronic band structures.{\cite{ZhigangZhang2018LNLTR3c,Hsu1997LT,QuentinJacquet2017Li3NbO4cubic,Boulay2003betaL3Treinvestigation}}

\subsection{Defect Formation Energies} 
\label{PyCDT}
The following intrinsic point defects were studied: Li vacancy ($\mathrm{Vac_{Li}}$) and M--Li antisite ($\mathrm{M_{Li}}$, M~=~Nb, Ta). We also investigated charge-neutral defect complexes $\left[4\mathrm{{Vac}_{Li}^{'}}+\mathrm{M_{Li}^{\bullet\bullet\bullet\bullet}}\right]$, M~=~Nb, Ta. Defects in \ce{LiNbO_3}, \ce{LiTaO_3}, \ce{Li_3NbO_4}, and \ce{Li_3TaO_4} were modeled by large supercells containing 320, 320, 256, and 256 atoms, respectively. The defect formation energies $\Delta E_D(X^q)$ are calculated with Eq.~\ref{DFEequation}:
\begin{equation}
\label{DFEequation}
\begin{aligned}
    \Delta E_D(X^q) & = E^{\mathrm{tot}}_{\mathrm{defect}} + E^{\mathrm{tot}}_{\mathrm{bulk}} +\\
    &- \underset{i}{\sum}n_i\mu_i + qE_{\ce{Fermi}} + E_{corr}.
\end{aligned}
\end{equation}
where \ce{E^{tot}_{defect}} and \ce{E^{tot}_{bulk}} are the r$^2$SCAN energies of supercells with and without defect. \ce{\mu_i} is the chemical potential of species $i$, while \ce{n_i} is the number of atoms of $i$ added to (${n_i}~>~0$) or removed from (${n_i}~<~0$) the supercell model.  $\mu_\mathrm{Li}$, and $\mu_\mathrm{M}$ were derived from the phase diagram of Li--Nb--O and Li--Ta--O (Figure S5). We explored defects with charges~$q$: $\left[-1,\ 0,\ \mathrm{and}\ 1\right]$ for $\mathrm{Vac_{Li}}$, $\left[-1,\ 0,\ 1,\ 2,\ 3,\ \mathrm{and}\ 4\right]$ for $\mathrm{M_{Li}}$. The \ce{E_{Fermi}} span the r$^2$SCAN band gaps of the respective bulk structures. \ce{E_{corr}} corrects for the electrostatic energy of charged defects interacting with their periodic images and the potential alignment for the fictitious jellium background.\cite{Freysold2009pycdt,Kumagai2014Obapycdtcorrection} The correction of Ref.~\citenum{Kumagai2014Obapycdtcorrection} was used for lower symmetry  \ce{LiNbO_3}, \ce{Li_3TaO_3}, and \ce{Li_3TaO_4},\cite{Kumagai2014Obapycdtcorrection} whereas we implemented Ref.~\citenum{Freysold2009pycdt} for \ce{Li_3NbO_4}. The band structures computed with different GGA (PBE and PBEsol) and meta-GGA (SCAN and r$^2$SCAN) exchange and correlation functionals are shown in Figure S4 of the SI. Details about the computation of defect concentrations and the identification of the equilibrium Fermi-energies are in Ref.s~\citenum{Gorai2021Defect,LiYuheng2022pointdefect}.

\subsection{Li-ion migration in \ce{\mathbf {LiMO_3}} and \ce{\mathbf{Li_3MO_4}} (M=Nb, Ta)} \label{NEB-M}
The empirical softBV analysis was used to identify the topology of Li-ion diffusion in \ce{LiMO_3} and \ce{Li_3MO_4}.\cite{adams2001relationship} 

In the nudged elastic band (NEB) calculations we used models as detailed in Section~\ref{PyCDT}, which ensure a minimum distance $\geq$8 {\AA} between image replicas.\cite{Henkelman2008NEB,TinaChen2019coatingthickness} In NEBs,  spring forces were set to 5~eV {\AA}$^{-1}$, and charge neutrality was imposed through  a compensating background charge. Models of neutral defect complexes with the lowest formation energies were selected from Section~\ref{PyCDT} used in NEBs.

To fit the moment tensor potentials (MTPs), the training sets were generated with \textit{ab initio} molecular dynamics (AIMD) simulations on $2\times2\times2$ supercells (256 atoms) of \ce{Li_3MO_4}, which are geometrically optimized by DFT using {\ce{r^2}SCAN} functional. An NVT ensemble based on the Nos\'{e}-Hoover thermostat, and a time step of 0.5~fs were applied in the AIMDs. \cite{Hoover1985thermostate} A plane wave energy cutoff of the AIMDs was 400~eV. The total energy was integrated at the $\Gamma$ point.\cite{Juefan2023MTPMDSE}

For \ce{Li_3NbO_4}, AIMDs we performed simulations at 600K, 800K, 1000K, and 1200K, each lasting 8~ps for the equilibration step, and after a temperature ramping of 0.5~ps. In total 8000 snapshots of the last 1~ps were taken as the training sets.\cite{Juefan2023MTPMDSE} While for \ce{Li_3TaO_4}, the temperature range was 600K--1500K with a step of 300K. We investigate the role of defects on the Li diffusion in \ce{Li_3MO_4}, by considering two types of defects, i.e., the Li vacancy (\ce{Vac_{Li}}) and the defect complex (4\ce{Vac_{Li}+Nb_{Li} (Ta_{Li})}). In the model of Li vacancy, a Li vacancy was introduced into a supercell of 256 atoms (a concentration of $\sim$1$\%$) and compensated with a background charge. We replaced one Li atom with an Nb (Ta) atom for complex antisite-vacancy defects and created four Li vacancies based on the most stable models calculated from Section~\ref{PyCDT}, with concentrations of antisites of $\sim$1$\%$ and Li vacancies of $\sim$4$\%$.

In training the MTPs, we chose the radius cutoff of 5~\AA, and a maximum level of 12 controlling the completeness of basis functions during the training of the bulk structures and the defective structures with 0.5\%, 1\%, 2\% Li vacancies.\cite{YunxingZuo2020MLIP,Juefan2023MTPMDSE} For structures with 4\% Li vacancies or with antisite defect complex, we increased $R_{cut}$ to 6 {\AA} to better capture the structural complexity of this situation. Weights on energies, forces, and stresses were set 100:10:1. The accuracy of the MTP fittings is reported in Table~S12.

MTPs molecular dynamics (MTP-MD) simulations were conducted in LAMMPS,\cite{Steve1994LAMMPS} using the NVT ensemble (Nos\'{e}--Hoover).\cite{Hoover1985thermostate}  We carried out long MDs of \ce{Li_3MO_4} for 10~ns with a timestep of 1~fs, preceded by a temperature ramping of 10~ps, followed by 1~ns of equilibration. MDs used 4$\times$4$\times$4 supercells with 2048 atoms. By varying the number of removed Li atoms, we computed the Li-ion conductivity of \ce{Li_3MO_4} structures including \ce{Vac_{Li}} concentrations of 0.5\%, 1\%, 2\%, and 4\%. By monitoring the mean-square displacements (see Figure S14 in the SI) of Li$^+$ in MTP-MDs, we extracted the tracer diffusivity of Eq.~\ref{eq:einstein}.
\begin{equation}
    D^* (T) = \lim_{t \rightarrow \infty}\frac{1}{2dt}\left(\frac{1}{N} \sum_{i=1} ^N\langle \left[ \vec{r}_i(t)-\vec{r}_i(0)\right]^2\rangle \right) 
    \label{eq:einstein}
\end{equation}
where $d=3$, $N$ is the number of lithium ions, $\vec{r} _i(t)$ is the displacement of $i^\mathrm{th}$ Li at time $t$. The activation energy of Li-ion migration ($E_a$) was derived from the Arrhenius equation. 
\begin{equation}
    D^* (T) = D_0 \, \exp\left(-\frac{E_a}{k_BT}\right)
\end{equation}
where $D_0$ is the prefactor, $T$ is the temperature, and $k_B$ is the Boltzmann constant. Li-ion conductivities $\sigma (T)$ are extracted from the Nernst--Einstein equation,
\begin{equation}
    \sigma (T) = \frac{n z^2 e^2}{k_BT}D^*(T)
\end{equation}
where $n$ is the volume density of the Li$^+$, $z~=~+1$, $e$ is the electron charge.

\begin{acknowledgement}
P.\ C.\ acknowledges funding from the National Research Foundation under its NRF Fellowship NRFF12-2020-0012. The computational work was performed on resources of the National Supercomputing Centre, Singapore (\url{https://www.nscc.sg}). 
\end{acknowledgement}

\begin{suppinfo}
The Supporting Information is available free of charge and includes (\emph{i}) The synthesis and applications of lithium niobate and tantalate coating materials. (\emph{ii}) The structures of lithium niobate and tantalate compounds. (\emph{iii}) The analysis of intrinsic defects in lithium niobate and tantalate compounds. (\emph{iv}) The Li-ion migration barriers and migration paths vs.\ niobate or tantalate local environments. (\emph{v}) The migration barriers extracted from moment-tensor potential molecular dynamics. All the computational data associated with this study are freely available at the repository \url{https://github.com/caneparesearch/LNOcoating-paper-data.git}.
\end{suppinfo}

\bibliography{biblio}

\providecommand{\latin}[1]{#1}
\makeatletter
\providecommand{\doi}
  {\begingroup\let\do\@makeother\dospecials
  \catcode`\{=1 \catcode`\}=2 \doi@aux}
\providecommand{\doi@aux}[1]{\endgroup\texttt{#1}}
\makeatother
\providecommand*\mcitethebibliography{\thebibliography}
\csname @ifundefined\endcsname{endmcitethebibliography}
  {\let\endmcitethebibliography\endthebibliography}{}
\begin{mcitethebibliography}{77}
\providecommand*\natexlab[1]{#1}
\providecommand*\mciteSetBstSublistMode[1]{}
\providecommand*\mciteSetBstMaxWidthForm[2]{}
\providecommand*\mciteBstWouldAddEndPuncttrue
  {\def\EndOfBibitem{\unskip.}}
\providecommand*\mciteBstWouldAddEndPunctfalse
  {\let\EndOfBibitem\relax}
\providecommand*\mciteSetBstMidEndSepPunct[3]{}
\providecommand*\mciteSetBstSublistLabelBeginEnd[3]{}
\providecommand*\EndOfBibitem{}
\mciteSetBstSublistMode{f}
\mciteSetBstMaxWidthForm{subitem}{(\alph{mcitesubitemcount})}
\mciteSetBstSublistLabelBeginEnd
  {\mcitemaxwidthsubitemform\space}
  {\relax}
  {\relax}

\bibitem[Li \latin{et~al.}(2021)Li, Duan, Li, Zhang, Deng, and
  Chen]{LiLiansheng2021LNcoatingreview}
Li,~L.; Duan,~H.; Li,~J.; Zhang,~L.; Deng,~Y.; Chen,~G. {Toward High
  Performance All-Solid-State Lithium Batteries with High-Voltage Cathode
  Materials: Design Strategies for Solid Electrolytes, Cathode Interfaces, and
  Composite Electrodes}. \emph{Adv. Energy Mater.} \textbf{2021}, \emph{11},
  2003154\relax
\mciteBstWouldAddEndPuncttrue
\mciteSetBstMidEndSepPunct{\mcitedefaultmidpunct}
{\mcitedefaultendpunct}{\mcitedefaultseppunct}\relax
\EndOfBibitem
\bibitem[Wang \latin{et~al.}(2015)Wang, Li, and
  Chen]{Wang2015interfacestabilityLIBs}
Wang,~K.~X.; Li,~X.~H.; Chen,~J.~S. {Surface and Interface Engineering of
  Electrode Materials for Lithium-Ion Batteries}. \emph{Adv. Mater.}
  \textbf{2015}, \emph{27}, 527--545\relax
\mciteBstWouldAddEndPuncttrue
\mciteSetBstMidEndSepPunct{\mcitedefaultmidpunct}
{\mcitedefaultendpunct}{\mcitedefaultseppunct}\relax
\EndOfBibitem
\bibitem[Zhang \latin{et~al.}(2017)Zhang, Schr\"{o}der, Arlt, Manke, Koerver,
  Pinedo, Weber, Sann, Zeier, and Janek]{WenboZhang2017expansion}
Zhang,~W.; Schr\"{o}der,~D.; Arlt,~T.; Manke,~I.; Koerver,~R.; Pinedo,~R.;
  Weber,~D.~A.; Sann,~J.; Zeier,~W.~G.; Janek,~J. (Electro)chemical expansion
  during cycling: monitoring the pressure changes in operating solid-state
  lithium batteries. \emph{J. Mater. Chem.A} \textbf{2017}, \emph{5},
  9929--9936\relax
\mciteBstWouldAddEndPuncttrue
\mciteSetBstMidEndSepPunct{\mcitedefaultmidpunct}
{\mcitedefaultendpunct}{\mcitedefaultseppunct}\relax
\EndOfBibitem
\bibitem[Famprikis \latin{et~al.}(2019)Famprikis, Canepa, Dawson, {Saiful
  Islam}, and Masquelier]{Piero2019SSEnature}
Famprikis,~T.; Canepa,~P.; Dawson,~J.~A.; {Saiful Islam},~M.; Masquelier,~C.
  {Fundamentals of inorganic solid-state electrolytes for batteries}.
  \emph{Nat. Mater.} \textbf{2019}, \emph{18}, 1278--1291\relax
\mciteBstWouldAddEndPuncttrue
\mciteSetBstMidEndSepPunct{\mcitedefaultmidpunct}
{\mcitedefaultendpunct}{\mcitedefaultseppunct}\relax
\EndOfBibitem
\bibitem[Tan \latin{et~al.}(2019)Tan, Wu, Nguyen, Chen, Marple, Doux, Wang,
  Yang, Banerjee, and Meng]{Tan2019thiophosphatestability}
Tan,~D. H.~S.; Wu,~E.~A.; Nguyen,~H.; Chen,~Z.; Marple,~M. A.~T.; Doux,~J.-M.;
  Wang,~X.; Yang,~H.; Banerjee,~A.; Meng,~Y.~S. {Elucidating Reversible
  Electrochemical Redox of Li$_{6}$PS$_{5}$Cl Solid Electrolyte}. \emph{ACS
  Energy Lett.} \textbf{2019}, \emph{4}, 2418--2427\relax
\mciteBstWouldAddEndPuncttrue
\mciteSetBstMidEndSepPunct{\mcitedefaultmidpunct}
{\mcitedefaultendpunct}{\mcitedefaultseppunct}\relax
\EndOfBibitem
\bibitem[Zhu \latin{et~al.}(2015)Zhu, He, and Mo]{YizhouZhu2015thermodynamicSE}
Zhu,~Y.; He,~X.; Mo,~Y. {Origin of Outstanding Stability in the Lithium Solid
  Electrolyte Materials: Insights from Thermodynamic Analyses Based on
  First-Principles Calculations}. \emph{ACS Appl. Mater. Interfaces}
  \textbf{2015}, \emph{7}, 23685--23693\relax
\mciteBstWouldAddEndPuncttrue
\mciteSetBstMidEndSepPunct{\mcitedefaultmidpunct}
{\mcitedefaultendpunct}{\mcitedefaultseppunct}\relax
\EndOfBibitem
\bibitem[Minnmann \latin{et~al.}(2022)Minnmann, Strauss, Bielefeld, Ruess,
  Adelhelm, Burkhardt, Dreyer, Trevisanello, Ehrenberg, Brezesinski, Richter,
  and Janek]{Minnmann2022SEcoating}
Minnmann,~P.; Strauss,~F.; Bielefeld,~A.; Ruess,~R.; Adelhelm,~P.;
  Burkhardt,~S.; Dreyer,~S.~L.; Trevisanello,~E.; Ehrenberg,~H.;
  Brezesinski,~T.; Richter,~F.~H.; Janek,~J. {Designing Cathodes and Cathode
  Active Materials for Solid-State Batteries}. \emph{Adv. Energy Mater.}
  \textbf{2022}, \emph{12}, 2201425\relax
\mciteBstWouldAddEndPuncttrue
\mciteSetBstMidEndSepPunct{\mcitedefaultmidpunct}
{\mcitedefaultendpunct}{\mcitedefaultseppunct}\relax
\EndOfBibitem
\bibitem[Chen \latin{et~al.}(2010)Chen, Qin, Amine, and
  Sun]{ZonghaiChen2010coatingLIBs}
Chen,~Z.; Qin,~Y.; Amine,~K.; Sun,~Y.-K. {Role of surface coating on cathode
  materials for lithium-ion batteries}. \emph{J. Mater. Chem.} \textbf{2010},
  \emph{20}, 7606--7612\relax
\mciteBstWouldAddEndPuncttrue
\mciteSetBstMidEndSepPunct{\mcitedefaultmidpunct}
{\mcitedefaultendpunct}{\mcitedefaultseppunct}\relax
\EndOfBibitem
\bibitem[Culver \latin{et~al.}(2019)Culver, Koerver, Zeier, and
  Janek]{Culver2019Funcionalitycoating}
Culver,~S.~P.; Koerver,~R.; Zeier,~W.~G.; Janek,~J. {On the Functionality of
  Coatings for Cathode Active Materials in Thiophosphate-Based All-Solid-State
  Batteries}. \emph{Adv. Energy Mater.} \textbf{2019}, \emph{9}, 1--14\relax
\mciteBstWouldAddEndPuncttrue
\mciteSetBstMidEndSepPunct{\mcitedefaultmidpunct}
{\mcitedefaultendpunct}{\mcitedefaultseppunct}\relax
\EndOfBibitem
\bibitem[Ohta \latin{et~al.}(2007)Ohta, Takada, Sakaguchi, Zhang, Ma, Fukuda,
  Osada, and Sasaki]{NarumiOhta2007}
Ohta,~N.; Takada,~K.; Sakaguchi,~I.; Zhang,~L.; Ma,~R.; Fukuda,~K.; Osada,~M.;
  Sasaki,~T. {LiNbO$_3$-coated LiCoO$_2$ as cathode material for all
  solid-state lithium secondary batteries}. \emph{Electrochem. commun.}
  \textbf{2007}, \emph{9}, 1486--1490\relax
\mciteBstWouldAddEndPuncttrue
\mciteSetBstMidEndSepPunct{\mcitedefaultmidpunct}
{\mcitedefaultendpunct}{\mcitedefaultseppunct}\relax
\EndOfBibitem
\bibitem[Peng \latin{et~al.}(2021)Peng, Ren, Zhang, Chen, Yu, Miao, Zhang, He,
  Yu, Zhang, Cheng, and Xie]{SSBNCM712LNO2021Peng}
Peng,~L.; Ren,~H.; Zhang,~J.; Chen,~S.; Yu,~C.; Miao,~X.; Zhang,~Z.; He,~Z.;
  Yu,~M.; Zhang,~L.; Cheng,~S.; Xie,~J. {LiNbO$_{3}$-coated
  LiNi$_{0.7}$Co$_{0.1}$Mn$_{0.2}$O$_{2}$ and chlorine-rich argyrodite enabling
  high-performance solid-state batteries under different temperatures}.
  \emph{Energy Storage Mater.} \textbf{2021}, \emph{43}, 53--61\relax
\mciteBstWouldAddEndPuncttrue
\mciteSetBstMidEndSepPunct{\mcitedefaultmidpunct}
{\mcitedefaultendpunct}{\mcitedefaultseppunct}\relax
\EndOfBibitem
\bibitem[Liu \latin{et~al.}(2020)Liu, Lu, Wan, Weng, Cai, Li, Que, Liu, and
  Yao]{GaozhanLiu2020LNcoating5VSSB}
Liu,~G.; Lu,~Y.; Wan,~H.; Weng,~W.; Cai,~L.; Li,~Z.; Que,~X.; Liu,~H.; Yao,~X.
  Passivation of the Cathode{\textendash}Electrolyte Interface for 5 V-Class
  All-Solid-State Batteries. \emph{ACS Appl. Mater. Interfaces} \textbf{2020},
  \emph{12}, 28083--28090\relax
\mciteBstWouldAddEndPuncttrue
\mciteSetBstMidEndSepPunct{\mcitedefaultmidpunct}
{\mcitedefaultendpunct}{\mcitedefaultseppunct}\relax
\EndOfBibitem
\bibitem[Heitjans \latin{et~al.}(2007)Heitjans, Masoud, Feldhoff, and
  Wilkening]{Heitjans2006NMRsingle}
Heitjans,~P.; Masoud,~M.; Feldhoff,~A.; Wilkening,~M. {NMR and impedance
  studies of nanocrystalline and amorphous ion conductors: lithium niobate as a
  model system}. \emph{Faraday Discuss.} \textbf{2007}, \emph{134},
  67--82\relax
\mciteBstWouldAddEndPuncttrue
\mciteSetBstMidEndSepPunct{\mcitedefaultmidpunct}
{\mcitedefaultendpunct}{\mcitedefaultseppunct}\relax
\EndOfBibitem
\bibitem[Wilkening \latin{et~al.}(2008)Wilkening, Epp, Feldhoff, and
  Heitjans]{Wilkening2008LTnanocrystalline}
Wilkening,~M.; Epp,~V.; Feldhoff,~A.; Heitjans,~P. {Tuning the Li Diffusivity
  of Poor Ionic Conductors by Mechanical Treatment: High Li Conductivity of
  Strongly Defective LiTaO$_{3}$ Nanoparticles}. \emph{J. Phys. Chem. C}
  \textbf{2008}, \emph{112}, 9291--9300\relax
\mciteBstWouldAddEndPuncttrue
\mciteSetBstMidEndSepPunct{\mcitedefaultmidpunct}
{\mcitedefaultendpunct}{\mcitedefaultseppunct}\relax
\EndOfBibitem
\bibitem[Zhang \latin{et~al.}(2014)Zhang, Zhou, Song, Shi, Wang, Guo, and
  Zhang]{YantaoZhang2014}
Zhang,~Y.; Zhou,~E.; Song,~D.; Shi,~X.; Wang,~X.; Guo,~J.; Zhang,~L. {Effects
  on electrochemical performances for host material caused by structure change
  of modifying material}. \emph{Phys. Chem. Chem. Phys} \textbf{2014},
  \emph{16}, 17792--17798\relax
\mciteBstWouldAddEndPuncttrue
\mciteSetBstMidEndSepPunct{\mcitedefaultmidpunct}
{\mcitedefaultendpunct}{\mcitedefaultseppunct}\relax
\EndOfBibitem
\bibitem[Lu \latin{et~al.}(2021)Lu, Peng, Zhang, Wang, Shi, Song, Zhang, and
  Zhang]{GuozhongLu2021Study}
Lu,~G.; Peng,~W.; Zhang,~Y.; Wang,~X.; Shi,~X.; Song,~D.; Zhang,~H.; Zhang,~L.
  {Study on the formation, development and coating mechanism of new phases on
  interface in LiNbO$_{3}$-coated LiCoO$_{2}$}. \emph{Electrochim. Acta}
  \textbf{2021}, \emph{368}, 137639\relax
\mciteBstWouldAddEndPuncttrue
\mciteSetBstMidEndSepPunct{\mcitedefaultmidpunct}
{\mcitedefaultendpunct}{\mcitedefaultseppunct}\relax
\EndOfBibitem
\bibitem[Xin \latin{et~al.}(2021)Xin, Zhou, Zong, Zuba, Chen, Chernova, Bai,
  Pei, Goel, Rana, Wang, An, Piper, Zhou, and
  Whittingham]{FengxiaXin2021WhatistheRole}
Xin,~F.; Zhou,~H.; Zong,~Y.; Zuba,~M.; Chen,~Y.; Chernova,~N.~A.; Bai,~J.;
  Pei,~B.; Goel,~A.; Rana,~J.; Wang,~F.; An,~K.; Piper,~L. F.~J.; Zhou,~G.;
  Whittingham,~M.~S. {What is the Role of Nb in Nickel-Rich Layered Oxide
  Cathodes for Lithium-Ion Batteries?} \emph{ACS Energy Lett.} \textbf{2021},
  \emph{6}, 1377--1382\relax
\mciteBstWouldAddEndPuncttrue
\mciteSetBstMidEndSepPunct{\mcitedefaultmidpunct}
{\mcitedefaultendpunct}{\mcitedefaultseppunct}\relax
\EndOfBibitem
\bibitem[Lee and Park(2021)Lee, and Park]{JunSuLee2021Comparison}
Lee,~J.~S.; Park,~Y.~J. {Comparison of LiTaO$_{3}$ and LiNbO$_{3}$ Surface
  Layers Prepared by Post- And Precursor-Based Coating Methods for Ni-Rich
  Cathodes of All-Solid-State Batteries}. \emph{ACS Appl. Mater. Interfaces}
  \textbf{2021}, \emph{13}, 38333--38345\relax
\mciteBstWouldAddEndPuncttrue
\mciteSetBstMidEndSepPunct{\mcitedefaultmidpunct}
{\mcitedefaultendpunct}{\mcitedefaultseppunct}\relax
\EndOfBibitem
\bibitem[Glass \latin{et~al.}(1978)Glass, Nassau, and Negran]{Glass1978Ionic}
Glass,~A.~M.; Nassau,~K.; Negran,~T.~J. {Ionic conductivity of quenched alkali
  niobate and tantalate glasses}. \emph{J. Appl. Phys.} \textbf{1978},
  \emph{49}, 4808--4811\relax
\mciteBstWouldAddEndPuncttrue
\mciteSetBstMidEndSepPunct{\mcitedefaultmidpunct}
{\mcitedefaultendpunct}{\mcitedefaultseppunct}\relax
\EndOfBibitem
\bibitem[Kim \latin{et~al.}(2020)Kim, Kim, Choi, and Park]{Kim2020Bifunctional}
Kim,~J.~H.; Kim,~H.; Choi,~W.; Park,~M.-S. {Bifunctional Surface Coating of
  LiNbO$_{3}$ on High-Ni Layered Cathode Materials for Lithium-Ion Batteries}.
  \emph{ACS Appl. Mater. Interfaces} \textbf{2020}, \emph{12},
  35098--35104\relax
\mciteBstWouldAddEndPuncttrue
\mciteSetBstMidEndSepPunct{\mcitedefaultmidpunct}
{\mcitedefaultendpunct}{\mcitedefaultseppunct}\relax
\EndOfBibitem
\bibitem[Lee \latin{et~al.}(2022)Lee, Hoang, Byeon, Jung, Moon, and
  Park]{HyoBinLee2022Surface}
Lee,~H.~B.; Hoang,~T.~D.; Byeon,~Y.~S.; Jung,~H.; Moon,~J.; Park,~M.-S. Surface
  Stabilization of Ni-Rich Layered Cathode Materials via Surface Engineering
  with LiTaO$_3$ for Lithium-Ion Batteries. \emph{ACS Appl. Mater. Interfaces}
  \textbf{2022}, \emph{14}, 2731--2741\relax
\mciteBstWouldAddEndPuncttrue
\mciteSetBstMidEndSepPunct{\mcitedefaultmidpunct}
{\mcitedefaultendpunct}{\mcitedefaultseppunct}\relax
\EndOfBibitem
\bibitem[Iyi \latin{et~al.}(1992)Iyi, Kitamura, Izumi, Yamamoto, Hayashi,
  Asano, and Kimura]{Iyi1992comparative}
Iyi,~N.; Kitamura,~K.; Izumi,~F.; Yamamoto,~J.~K.; Hayashi,~T.; Asano,~H.;
  Kimura,~S. {Comparative Study of Defect Structures in Lithium Niobate with
  Different Compositions}. \emph{J. Solid State Chem.} \textbf{1992},
  \emph{101}, 340--343\relax
\mciteBstWouldAddEndPuncttrue
\mciteSetBstMidEndSepPunct{\mcitedefaultmidpunct}
{\mcitedefaultendpunct}{\mcitedefaultseppunct}\relax
\EndOfBibitem
\bibitem[Smyth(1983)]{Smyth1983LNdefect}
Smyth,~D.~M. {Defects and transport in LiNbO$_{3}$}. \emph{Ferroelectrics}
  \textbf{1983}, \emph{50}, 93--102\relax
\mciteBstWouldAddEndPuncttrue
\mciteSetBstMidEndSepPunct{\mcitedefaultmidpunct}
{\mcitedefaultendpunct}{\mcitedefaultseppunct}\relax
\EndOfBibitem
\bibitem[Kolb and Laudise(1976)Kolb, and Laudise]{Kolb1976pdLTO}
Kolb,~E.~D.; Laudise,~R.~A. {The phase diagram, LiOH-Ta$_2$O$_5$-H$_2$O and the
  hydrothermal synthesis of LiTaO$_3$ and LiNbO$_3$}. \emph{J. Cryst. Growth}
  \textbf{1976}, \emph{33}, 145--149\relax
\mciteBstWouldAddEndPuncttrue
\mciteSetBstMidEndSepPunct{\mcitedefaultmidpunct}
{\mcitedefaultendpunct}{\mcitedefaultseppunct}\relax
\EndOfBibitem
\bibitem[Bernasconi \latin{et~al.}(1999)Bernasconi, Montemezzani, G{\"{u}}nter,
  Furukawa, and Kitamura]{Bernasconi1999LTantisite}
Bernasconi,~P.; Montemezzani,~G.; G{\"{u}}nter,~P.; Furukawa,~Y.; Kitamura,~K.
  {Stoichiometric LiTaO$_3$ for ultraviolet photorefraction}.
  \emph{Ferroelectrics} \textbf{1999}, \emph{223}, 373--379\relax
\mciteBstWouldAddEndPuncttrue
\mciteSetBstMidEndSepPunct{\mcitedefaultmidpunct}
{\mcitedefaultendpunct}{\mcitedefaultseppunct}\relax
\EndOfBibitem
\bibitem[Rahn \latin{et~al.}(2012)Rahn, H{\"{u}}ger, D{\"{o}}rrer, Ruprecht,
  Heitjans, and Schmidt]{Rahn2011congruentLNconductivity}
Rahn,~J.; H{\"{u}}ger,~E.; D{\"{o}}rrer,~L.; Ruprecht,~B.; Heitjans,~P.;
  Schmidt,~H. {Li self-diffusion in lithium niobate single crystals at low
  temperatures}. \emph{Phys. Chem. Chem. Phys.} \textbf{2012}, \emph{14},
  2427--2433\relax
\mciteBstWouldAddEndPuncttrue
\mciteSetBstMidEndSepPunct{\mcitedefaultmidpunct}
{\mcitedefaultendpunct}{\mcitedefaultseppunct}\relax
\EndOfBibitem
\bibitem[H{\"{u}}ger \latin{et~al.}(2014)H{\"{u}}ger, Rahn, Stahn, Geue,
  {Heitjans Cd}, and Schmidt]{Huger2014singleLiNbO3Ea}
H{\"{u}}ger,~E.; Rahn,~J.; Stahn,~J.; Geue,~T.; {Heitjans Cd},~P.; Schmidt,~H.
  {Lithium diffusion in congruent LiNbO$_3$ single crystals at low temperatures
  probed by neutron reflectometry}. \emph{Phys. Chem. Chem. Phys}
  \textbf{2014}, \emph{16}, 3670\relax
\mciteBstWouldAddEndPuncttrue
\mciteSetBstMidEndSepPunct{\mcitedefaultmidpunct}
{\mcitedefaultendpunct}{\mcitedefaultseppunct}\relax
\EndOfBibitem
\bibitem[Xin \latin{et~al.}(2021)Xin, Zhou, Bai, Wang, and
  Whittingham]{Xin2021Conditioning}
Xin,~F.; Zhou,~H.; Bai,~J.; Wang,~F.; Whittingham,~M.~S. {Conditioning the
  Surface and Bulk of High-Nickel Cathodes with a Nb Coating: An In Situ X-ray
  Study}. \emph{J. Phys. Chem. Lett.} \textbf{2021}, \emph{12},
  7908--7913\relax
\mciteBstWouldAddEndPuncttrue
\mciteSetBstMidEndSepPunct{\mcitedefaultmidpunct}
{\mcitedefaultendpunct}{\mcitedefaultseppunct}\relax
\EndOfBibitem
\bibitem[Liao \latin{et~al.}(2013)Liao, Singh, Li, and
  Goodenough]{Liao2013L3NO4barrierWsubstitution}
Liao,~Y.; Singh,~P.; Li,~W.; Goodenough,~J.~B. {Comparison of Li$^+$
  conductivity in Li$_{3-x}$Nb$_{1-x}$M$_x$O$_4$ (M = W, Mo) with that in
  Li$_{3-2x}$Ni$_x$NbO$_4$}. \emph{Mater. Res. Bull.} \textbf{2013}, \emph{48},
  1372--1375\relax
\mciteBstWouldAddEndPuncttrue
\mciteSetBstMidEndSepPunct{\mcitedefaultmidpunct}
{\mcitedefaultendpunct}{\mcitedefaultseppunct}\relax
\EndOfBibitem
\bibitem[Yabuuchi \latin{et~al.}(2015)Yabuuchi, Takeuchi, Nakayama, Shiiba,
  Ogawa, Nakayama, Ohta, Endo, Ozaki, Inamasu, Sato, and
  Komaba]{Yabuuchi2015Li3NbO4electrode}
Yabuuchi,~N.; Takeuchi,~M.; Nakayama,~M.; Shiiba,~H.; Ogawa,~M.; Nakayama,~K.;
  Ohta,~T.; Endo,~D.; Ozaki,~T.; Inamasu,~T.; Sato,~K.; Komaba,~S.
  {High-capacity electrode materials for rechargeable lithium batteries:
  Li$_3$NbO$_4$-based system with cation-disordered rocksalt structure}.
  \emph{PNAS} \textbf{2015}, \emph{112}, 7650--7655\relax
\mciteBstWouldAddEndPuncttrue
\mciteSetBstMidEndSepPunct{\mcitedefaultmidpunct}
{\mcitedefaultendpunct}{\mcitedefaultseppunct}\relax
\EndOfBibitem
\bibitem[Adams(2006)]{adams2006bond}
Adams,~S. {From bond valence maps to energy landscapes for mobile ions in
  ion-conducting solids}. \emph{Solid State Ionics} \textbf{2006}, \emph{177},
  1625--1630\relax
\mciteBstWouldAddEndPuncttrue
\mciteSetBstMidEndSepPunct{\mcitedefaultmidpunct}
{\mcitedefaultendpunct}{\mcitedefaultseppunct}\relax
\EndOfBibitem
\bibitem[Abrahams \latin{et~al.}(1966)Abrahams, Reddy, and
  Bernstein]{Abrahams1966LNR3c}
Abrahams,~S.~C.; Reddy,~J.~M.; Bernstein,~J.~L. {Ferroelectric lithium niobate
  3. Single crystal X-ray diffraction study at 24°C}. \emph{J. Phys. Chem.
  Solids Pergamon Press} \textbf{1966}, \emph{27}, 997--1012\relax
\mciteBstWouldAddEndPuncttrue
\mciteSetBstMidEndSepPunct{\mcitedefaultmidpunct}
{\mcitedefaultendpunct}{\mcitedefaultseppunct}\relax
\EndOfBibitem
\bibitem[Volk and W{\"{o}}hlecke(2008)Volk, and
  W{\"{o}}hlecke]{Volk2008LNphasediagram}
Volk,~T.; W{\"{o}}hlecke,~M. \emph{{Lithium Niobate: Defects, Photorefraction
  and Ferroelectric Switching}}; Springer Series in Materials Science vol 115:
  Berlin: Springer, 2008; p~4\relax
\mciteBstWouldAddEndPuncttrue
\mciteSetBstMidEndSepPunct{\mcitedefaultmidpunct}
{\mcitedefaultendpunct}{\mcitedefaultseppunct}\relax
\EndOfBibitem
\bibitem[Nyman \latin{et~al.}(2009)Nyman, Anderson, and
  Provencio]{MayNyman2009L3NL3Tsynthesis}
Nyman,~M.; Anderson,~T.~M.; Provencio,~P.~P. {Comparison of Aqueous and
  Non-aqueous Soft-Chemical Syntheses of Lithium Niobate and Lithium Tantalate
  Powders}. \emph{Cryst. Growth Des.} \textbf{2009}, \emph{9}, 1036--1040\relax
\mciteBstWouldAddEndPuncttrue
\mciteSetBstMidEndSepPunct{\mcitedefaultmidpunct}
{\mcitedefaultendpunct}{\mcitedefaultseppunct}\relax
\EndOfBibitem
\bibitem[Grenier \latin{et~al.}(1964)Grenier, Martin, and
  Durif-Varambon]{Grenier1964Fm3mLNLT}
Grenier,~J.-C.; Martin,~C.; Durif-Varambon,~A. {{\'{E}}tude cristallographique
  des orthoniobates et orthotantalates de lithium}. \emph{Bull.
  Min{\'{e}}ralogie} \textbf{1964}, \emph{87}, 316--320\relax
\mciteBstWouldAddEndPuncttrue
\mciteSetBstMidEndSepPunct{\mcitedefaultmidpunct}
{\mcitedefaultendpunct}{\mcitedefaultseppunct}\relax
\EndOfBibitem
\bibitem[Hsiao \latin{et~al.}(2010)Hsiao, Fang, Lin, Shieh, and
  Ji]{Hsiao2010L3Nbandgap}
Hsiao,~Y.-J.; Fang,~T.-H.; Lin,~S.-J.; Shieh,~J.-M.; Ji,~L.-W. {Preparation and
  luminescent characteristic of Li$_3$NbO$_4$ nanophosphor}. \emph{J. Lumin.}
  \textbf{2010}, \emph{130}, 1863--1865\relax
\mciteBstWouldAddEndPuncttrue
\mciteSetBstMidEndSepPunct{\mcitedefaultmidpunct}
{\mcitedefaultendpunct}{\mcitedefaultseppunct}\relax
\EndOfBibitem
\bibitem[Modeshia \latin{et~al.}(2010)Modeshia, Walton, Mitchell, and
  E.Ashbrook]{Deena2010Li3NO4disorderedtransform}
Modeshia,~D.~R.; Walton,~R.~I.; Mitchell,~M.~R.; E.Ashbrook,~S. {Disordered
  lithium niobate rock-salt materials prepared by hydrothermal synthesis}.
  \emph{Dalt. Trans.} \textbf{2010}, \emph{39}, 6021--6036\relax
\mciteBstWouldAddEndPuncttrue
\mciteSetBstMidEndSepPunct{\mcitedefaultmidpunct}
{\mcitedefaultendpunct}{\mcitedefaultseppunct}\relax
\EndOfBibitem
\bibitem[Mather \latin{et~al.}(2000)Mather, Dussarrat, Etourneau, and
  West]{Mather2000Li3NbO4structure}
Mather,~G.~C.; Dussarrat,~C.; Etourneau,~J.; West,~A.~R. {A review of
  cation-ordered rock salt superstructure oxides}. \emph{J. Mater. Chem.}
  \textbf{2000}, \emph{10}, 2219--2230\relax
\mciteBstWouldAddEndPuncttrue
\mciteSetBstMidEndSepPunct{\mcitedefaultmidpunct}
{\mcitedefaultendpunct}{\mcitedefaultseppunct}\relax
\EndOfBibitem
\bibitem[Zocchi \latin{et~al.}(1983)Zocchi, Gatt1, Santoro, and
  Roth]{Zocchi1983Li3TaO4}
Zocchi,~M.; Gatt1,~M.; Santoro,~A.; Roth,~R.~S. {Neutron and X-Ray Diffraction
  Study on Polymorphism in Lithium Orthotantalate, Li$_3$TaO$_4$}. \emph{J.
  Solid State Chem.} \textbf{1983}, \emph{48}, 420--430\relax
\mciteBstWouldAddEndPuncttrue
\mciteSetBstMidEndSepPunct{\mcitedefaultmidpunct}
{\mcitedefaultendpunct}{\mcitedefaultseppunct}\relax
\EndOfBibitem
\bibitem[{Du Boulay} \latin{et~al.}(2003){Du Boulay}, Sakaguchi, Suda, and
  Ishizawa]{Boulay2003betaL3Treinvestigation}
{Du Boulay},~D.; Sakaguchi,~A.; Suda,~K.; Ishizawa,~N. {Reinvestigation
  $\beta$-Li$_3$TaO$_4$}. \emph{Acta Crystallogr. Sect. E Struct. Reports}
  \textbf{2003}, \emph{59}, i80--i82\relax
\mciteBstWouldAddEndPuncttrue
\mciteSetBstMidEndSepPunct{\mcitedefaultmidpunct}
{\mcitedefaultendpunct}{\mcitedefaultseppunct}\relax
\EndOfBibitem
\bibitem[Kim \latin{et~al.}(2022)Kim, Pham, Lee, and
  Kim]{ChaeeunKim2022Li3TaO4polymorphs}
Kim,~C.; Pham,~T.~L.; Lee,~J.-S.; Kim,~Y.-I. {Synthesis, thermal analysis, and
  band gap of ordered and disordered complex rock salt Li$_3$TaO$_4$}. \emph{J.
  Solid State Chem.} \textbf{2022}, \emph{315}, 123450\relax
\mciteBstWouldAddEndPuncttrue
\mciteSetBstMidEndSepPunct{\mcitedefaultmidpunct}
{\mcitedefaultendpunct}{\mcitedefaultseppunct}\relax
\EndOfBibitem
\bibitem[Allemann \latin{et~al.}(1996)Allemann, Xia, Morriss, Wilkinson,
  Eckert, Speck, Levi, Lange, and Anderson]{Allemann1996LTphasediagram}
Allemann,~J.; Xia,~Y.; Morriss,~R.~E.; Wilkinson,~A.~P.; Eckert,~H.; Speck,~J.;
  Levi,~C.~G.; Lange,~F.~F.; Anderson,~S. Crystallization behavior of
  Li$_{1-5x}$Ta$_{1+x}$O$_3$ glasses synthesized from liquid precursors.
  \emph{Journal of Materials Research} \textbf{1996}, \emph{11},
  2376--2387\relax
\mciteBstWouldAddEndPuncttrue
\mciteSetBstMidEndSepPunct{\mcitedefaultmidpunct}
{\mcitedefaultendpunct}{\mcitedefaultseppunct}\relax
\EndOfBibitem
\bibitem[Xiao \latin{et~al.}(2019)Xiao, Miara, Wang, and
  Ceder]{YihanXiao2019computationalcoatingscreen}
Xiao,~Y.; Miara,~L.~J.; Wang,~Y.; Ceder,~G. {Computational Screening of Cathode
  Coatings for Solid-State Batteries}. \emph{Joule} \textbf{2019}, \emph{3},
  1252--1275\relax
\mciteBstWouldAddEndPuncttrue
\mciteSetBstMidEndSepPunct{\mcitedefaultmidpunct}
{\mcitedefaultendpunct}{\mcitedefaultseppunct}\relax
\EndOfBibitem
\bibitem[Zhu \latin{et~al.}(2016)Zhu, He, and
  Mo]{YizhouZhu2016Firstprinciplecoating}
Zhu,~Y.; He,~X.; Mo,~Y. {First principles study on electrochemical and chemical
  stability of solid electrolyte-electrode interfaces in all-solid-state Li-ion
  batteries}. \emph{J. Mater. Chem. A} \textbf{2016}, \emph{4}, 3253\relax
\mciteBstWouldAddEndPuncttrue
\mciteSetBstMidEndSepPunct{\mcitedefaultmidpunct}
{\mcitedefaultendpunct}{\mcitedefaultseppunct}\relax
\EndOfBibitem
\bibitem[SCOTT and BURNS(1972)SCOTT, and BURNS]{Scott1972expstoichiometryLNLT}
SCOTT,~B.~A.; BURNS,~G. {Determination of Stoichiometry Variations in LiNbO$_3$
  and LiTaO$_3$ by Raman Powder Spectroscopy}. \emph{J. Am. Ceram. Soc.}
  \textbf{1972}, \emph{55}, 225--230\relax
\mciteBstWouldAddEndPuncttrue
\mciteSetBstMidEndSepPunct{\mcitedefaultmidpunct}
{\mcitedefaultendpunct}{\mcitedefaultseppunct}\relax
\EndOfBibitem
\bibitem[Lerner \latin{et~al.}(1968)Lerner, Legras, and Dumas]{Lerner1968}
Lerner,~P.; Legras,~C.; Dumas,~J.~P. {Stoichiometry of lithium metaniobate
  single crystals}. \emph{J. Cryst. Growth} \textbf{1968}, \emph{3},
  231--235\relax
\mciteBstWouldAddEndPuncttrue
\mciteSetBstMidEndSepPunct{\mcitedefaultmidpunct}
{\mcitedefaultendpunct}{\mcitedefaultseppunct}\relax
\EndOfBibitem
\bibitem[Wilkinson \latin{et~al.}(1993)Wilkinson, Cheetham, and
  Jarman]{Wilkinson1993defectLiNbO3}
Wilkinson,~A.~P.; Cheetham,~A.~K.; Jarman,~R.~H. {The defect structure of
  congruently melting lithium niobate}. \emph{J. Appl. Phys.} \textbf{1993},
  \emph{74}, 3080--3083\relax
\mciteBstWouldAddEndPuncttrue
\mciteSetBstMidEndSepPunct{\mcitedefaultmidpunct}
{\mcitedefaultendpunct}{\mcitedefaultseppunct}\relax
\EndOfBibitem
\bibitem[Bl{\"{u}}mel \latin{et~al.}(1994)Bl{\"{u}}mel, Born, and
  Metzger]{Blumel1994ssNMRLN}
Bl{\"{u}}mel,~J.; Born,~E.; Metzger,~T. {Solid state NMR study supporting the
  lithium vacancy defect model in congruent lithium niobate}. \emph{J. Phys.
  Chem. Solids} \textbf{1994}, \emph{55}, 589--593\relax
\mciteBstWouldAddEndPuncttrue
\mciteSetBstMidEndSepPunct{\mcitedefaultmidpunct}
{\mcitedefaultendpunct}{\mcitedefaultseppunct}\relax
\EndOfBibitem
\bibitem[Xu \latin{et~al.}(2008)Xu, Lee, He, Sinnott, Gopalan, Dierolf, and
  Phillpot]{HaixuanXu2008Stability}
Xu,~H.; Lee,~D.; He,~J.; Sinnott,~S.~B.; Gopalan,~V.; Dierolf,~V.;
  Phillpot,~S.~R. {Stability of intrinsic defects and defect clusters in
  LiNbO$_3$ from density functional theory calculations}. \emph{Phys. Rev. B}
  \textbf{2008}, \emph{78}, 174103\relax
\mciteBstWouldAddEndPuncttrue
\mciteSetBstMidEndSepPunct{\mcitedefaultmidpunct}
{\mcitedefaultendpunct}{\mcitedefaultseppunct}\relax
\EndOfBibitem
\bibitem[Donnerberg \latin{et~al.}(1989)Donnerberg, Tomlinson, Catlow, and
  Schirmer]{Donnerberg1989}
Donnerberg,~H.; Tomlinson,~S.~M.; Catlow,~C. R.~A.; Schirmer,~O.~F.
  {Computer-simulation studies of intrinsic defects in LiNbO$_3$ crystals}.
  \emph{Phys. Rev. B} \textbf{1989}, \emph{40}, 11909--11916\relax
\mciteBstWouldAddEndPuncttrue
\mciteSetBstMidEndSepPunct{\mcitedefaultmidpunct}
{\mcitedefaultendpunct}{\mcitedefaultseppunct}\relax
\EndOfBibitem
\bibitem[Kuganathan \latin{et~al.}(2019)Kuganathan, Kordatos, Kelaidis, and
  Chroneos]{Kuganathan2019Li3NbO4defect}
Kuganathan,~N.; Kordatos,~A.; Kelaidis,~N.; Chroneos,~A. {Defects, Lithium
  Mobility and Tetravalent Dopants in the Li$_3$NbO$_4$ Cathode Material}.
  \emph{Sci. Rep.} \textbf{2019}, \emph{9}, 2192\relax
\mciteBstWouldAddEndPuncttrue
\mciteSetBstMidEndSepPunct{\mcitedefaultmidpunct}
{\mcitedefaultendpunct}{\mcitedefaultseppunct}\relax
\EndOfBibitem
\bibitem[Canepa \latin{et~al.}(2017)Canepa, Gopalakrishnan, Broberg, Bo, and
  Ceder]{Piero2017pointdefect}
Canepa,~P.; Gopalakrishnan,~S.,~Gautam; Broberg,~D.; Bo,~S.-H.; Ceder,~G. {Role
  of Point Defects in Spinel Mg Chalcogenide Conductors}. \emph{Chem. Mater.}
  \textbf{2017}, \emph{29}, 9657--9667\relax
\mciteBstWouldAddEndPuncttrue
\mciteSetBstMidEndSepPunct{\mcitedefaultmidpunct}
{\mcitedefaultendpunct}{\mcitedefaultseppunct}\relax
\EndOfBibitem
\bibitem[Li \latin{et~al.}(2022)Li, Canepa, and Gorai]{LiYuheng2022pointdefect}
Li,~Y.; Canepa,~P.; Gorai,~P. {Role of Electronic Passivation in Stabilizing
  the Lithium-Li$_x$PO$_y$N$_z$ Solid-Electrolyte Interphase}. \emph{PRX
  ENERGY} \textbf{2022}, \emph{1}, 023004\relax
\mciteBstWouldAddEndPuncttrue
\mciteSetBstMidEndSepPunct{\mcitedefaultmidpunct}
{\mcitedefaultendpunct}{\mcitedefaultseppunct}\relax
\EndOfBibitem
\bibitem[Li \latin{et~al.}(2015)Li, Schmidt, and Sanna]{YanluLi2015Defect}
Li,~Y.; Schmidt,~W.~G.; Sanna,~S. {Defect complexes in congruent LiNbO$_3$ and
  their optical signatures}. \emph{Phys. Rev. B} \textbf{2015}, \emph{91},
  174106\relax
\mciteBstWouldAddEndPuncttrue
\mciteSetBstMidEndSepPunct{\mcitedefaultmidpunct}
{\mcitedefaultendpunct}{\mcitedefaultseppunct}\relax
\EndOfBibitem
\bibitem[Gubaev \latin{et~al.}(2019)Gubaev, Podryabinkin, Hart, and
  Shapeev]{MTP2019KonstantinGubaev}
Gubaev,~K.; Podryabinkin,~E.~V.; Hart,~G. L.~W.; Shapeev,~A.~V. {Accelerating
  high-throughput searches for new alloys with active learning of interatomic
  potentials}. \emph{Comput. Mater. Sci.} \textbf{2019}, \emph{156},
  148--156\relax
\mciteBstWouldAddEndPuncttrue
\mciteSetBstMidEndSepPunct{\mcitedefaultmidpunct}
{\mcitedefaultendpunct}{\mcitedefaultseppunct}\relax
\EndOfBibitem
\bibitem[Ruprecht and Heitjans(2010)Ruprecht, and
  Heitjans]{2010Li3NbO4NMRultraslow}
Ruprecht,~B.; Heitjans,~P. {Ultraslow lithium diffusion in Li$_3$NbO$_4$ probed
  by 7Li simulated Echo NMR Spectroscopy}. \emph{Diffus. Fundam.}
  \textbf{2010}, \emph{12}, 100--101\relax
\mciteBstWouldAddEndPuncttrue
\mciteSetBstMidEndSepPunct{\mcitedefaultmidpunct}
{\mcitedefaultendpunct}{\mcitedefaultseppunct}\relax
\EndOfBibitem
\bibitem[Kim and Kim(2023)Kim, and Kim]{ChaeeunK2023L3TaO4defectconductivity}
Kim,~C.; Kim,~Y.-I. {Ionic Conductivity of Li$_3$TaO$_4$ Depending on
  Polymorphism and Non-Stoichiometric Defects}. \emph{Soc. Sci. Res. Netw.}
  \textbf{2023}, \relax
\mciteBstWouldAddEndPunctfalse
\mciteSetBstMidEndSepPunct{\mcitedefaultmidpunct}
{}{\mcitedefaultseppunct}\relax
\EndOfBibitem
\bibitem[Xin \latin{et~al.}(2022)Xin, Goel, Chen, Zhou, Bai, Liu, Wang, Zhou,
  and Whittingham]{XinFengxia2022Nbmechanical}
Xin,~F.; Goel,~A.; Chen,~X.; Zhou,~H.; Bai,~J.; Liu,~S.; Wang,~F.; Zhou,~G.;
  Whittingham,~M.~S. {Electrochemical Characterization and Microstructure
  Evolution of Ni-Rich Layered Cathode Materials by Niobium
  Coating/Substitution}. \emph{Chem. Mater.} \textbf{2022}, \emph{34},
  7858--7866\relax
\mciteBstWouldAddEndPuncttrue
\mciteSetBstMidEndSepPunct{\mcitedefaultmidpunct}
{\mcitedefaultendpunct}{\mcitedefaultseppunct}\relax
\EndOfBibitem
\bibitem[Chen \latin{et~al.}(2019)Chen, {Sai Gautam}, and
  Canepa]{TinaChen2019coatingthickness}
Chen,~T.; {Sai Gautam},~G.; Canepa,~P. {Ionic Transport in Potential Coating
  Materials for Mg Batteries}. \emph{Chem. Mater.} \textbf{2019}, \emph{31},
  8087--8099\relax
\mciteBstWouldAddEndPuncttrue
\mciteSetBstMidEndSepPunct{\mcitedefaultmidpunct}
{\mcitedefaultendpunct}{\mcitedefaultseppunct}\relax
\EndOfBibitem
\bibitem[Gorai \latin{et~al.}(2021)Gorai, Famprikis, Singh, Stevanovi{\'{c}},
  and Canepa]{Gorai2021Defect}
Gorai,~P.; Famprikis,~T.; Singh,~B.; Stevanovi{\'{c}},~V.; Canepa,~P. Devil is
  in the Defects: Electronic Conductivity in Solid Electrolytes. \emph{Chem.
  Mater.} \textbf{2021}, \emph{33}, 7484--7498\relax
\mciteBstWouldAddEndPuncttrue
\mciteSetBstMidEndSepPunct{\mcitedefaultmidpunct}
{\mcitedefaultendpunct}{\mcitedefaultseppunct}\relax
\EndOfBibitem
\bibitem[Kresse and Furthm{\"{u}}(1996)Kresse, and
  Furthm{\"{u}}]{Kresse1996VASPCondens}
Kresse,~G.; Furthm{\"{u}},~J. {Efficient iterative schemes for ab initio
  total-energy calculations using a plane-wave basis set}. \emph{Phys. Rev. B}
  \textbf{1996}, \emph{54}, 11169--11186\relax
\mciteBstWouldAddEndPuncttrue
\mciteSetBstMidEndSepPunct{\mcitedefaultmidpunct}
{\mcitedefaultendpunct}{\mcitedefaultseppunct}\relax
\EndOfBibitem
\bibitem[Kresse and Joubert(1999)Kresse, and Joubert]{Kresse1999PAWpotential}
Kresse,~G.; Joubert,~D. {From ultrasoft pseudopotentials to the projector
  augmented-wave method}. \emph{Phys. Rev. B} \textbf{1999}, \emph{59},
  1758--1775\relax
\mciteBstWouldAddEndPuncttrue
\mciteSetBstMidEndSepPunct{\mcitedefaultmidpunct}
{\mcitedefaultendpunct}{\mcitedefaultseppunct}\relax
\EndOfBibitem
\bibitem[Furness \latin{et~al.}(2020)Furness, Kaplan, Ning, Perdew, and
  Sun]{Furness2020R2scan}
Furness,~J.~W.; Kaplan,~A.~D.; Ning,~J.; Perdew,~J.~P.; Sun,~J. {Accurate and
  Numerically Efficient r$^2$SCAN Meta-Generalized Gradient Approximation}.
  \emph{J. Phys. Chem. Lett.} \textbf{2020}, \emph{11}, 8208--8215\relax
\mciteBstWouldAddEndPuncttrue
\mciteSetBstMidEndSepPunct{\mcitedefaultmidpunct}
{\mcitedefaultendpunct}{\mcitedefaultseppunct}\relax
\EndOfBibitem
\bibitem[Perdew \latin{et~al.}(1996)Perdew, Burke, and
  Ernzerhof]{Perdew1996PBE}
Perdew,~J.~P.; Burke,~K.; Ernzerhof,~M. Generalized Gradient Approximation Made
  Simple. \emph{Physical Review Letters} \textbf{1996}, \emph{77},
  3865--3868\relax
\mciteBstWouldAddEndPuncttrue
\mciteSetBstMidEndSepPunct{\mcitedefaultmidpunct}
{\mcitedefaultendpunct}{\mcitedefaultseppunct}\relax
\EndOfBibitem
\bibitem[{Ping Ong} \latin{et~al.}(2013){Ping Ong}, {Davidson Richards}, Jain,
  Hautier, Kocher, Cholia, Gunter, Chevrier, Persson, and
  Ceder]{ShyuePingOng2013pymatgen}
{Ping Ong},~S.; {Davidson Richards},~W.; Jain,~A.; Hautier,~G.; Kocher,~M.;
  Cholia,~S.; Gunter,~D.; Chevrier,~V.~L.; Persson,~K.~A.; Ceder,~G. {Python
  Materials Genomics (pymatgen): A robust, open-source python library for
  materials analysis}. \emph{Comput. Mater. Sci.} \textbf{2013}, \emph{68},
  314--319\relax
\mciteBstWouldAddEndPuncttrue
\mciteSetBstMidEndSepPunct{\mcitedefaultmidpunct}
{\mcitedefaultendpunct}{\mcitedefaultseppunct}\relax
\EndOfBibitem
\bibitem[Zhang \latin{et~al.}(2018)Zhang, Abe, Moriyoshi, Tanaka, and
  Kuroiwa]{ZhigangZhang2018LNLTR3c}
Zhang,~Z.~G.; Abe,~T.; Moriyoshi,~C.; Tanaka,~H.; Kuroiwa,~Y. {Study of
  materials structure physics of isomorphic LiNbO$_3$ and LiTaO$_3$
  ferroelectrics by synchrotron radiation X-ray diffraction}. \emph{Jpn. J.
  Appl. Phys.} \textbf{2018}, \emph{57}, 11UB04\relax
\mciteBstWouldAddEndPuncttrue
\mciteSetBstMidEndSepPunct{\mcitedefaultmidpunct}
{\mcitedefaultendpunct}{\mcitedefaultseppunct}\relax
\EndOfBibitem
\bibitem[Hsu \latin{et~al.}(1997)Hsu, Maslen, du~Boulay, and
  Ishizawa]{Hsu1997LT}
Hsu,~R.; Maslen,~E.~N.; du~Boulay,~D.; Ishizawa,~N. Synchrotron X-ray Studies
  of LiNbO$_3$ and LiTaO$_3$. \emph{Acta Crystallogr. Sect. B Struct. Sci.}
  \textbf{1997}, \emph{53}, 420--428\relax
\mciteBstWouldAddEndPuncttrue
\mciteSetBstMidEndSepPunct{\mcitedefaultmidpunct}
{\mcitedefaultendpunct}{\mcitedefaultseppunct}\relax
\EndOfBibitem
\bibitem[Jacquet \latin{et~al.}(2017)Jacquet, Perez, Batuk, Tendeloo, Rousse,
  and Tarascon]{QuentinJacquet2017Li3NbO4cubic}
Jacquet,~Q.; Perez,~A.; Batuk,~D.; Tendeloo,~G.~V.; Rousse,~G.; Tarascon,~J.-M.
  The Li$_3$Ru$_y$Nb$_{1-y}$O$_4$ (0$\leq$y$\leq$1) System: Structural
  Diversity and Li Insertion and Extraction Capabilities. \emph{Chemistry of
  Materials} \textbf{2017}, \emph{29}, 5331--5343\relax
\mciteBstWouldAddEndPuncttrue
\mciteSetBstMidEndSepPunct{\mcitedefaultmidpunct}
{\mcitedefaultendpunct}{\mcitedefaultseppunct}\relax
\EndOfBibitem
\bibitem[Freysoldt \latin{et~al.}(2009)Freysoldt, Neugebauer, and {Van De
  Walle}]{Freysold2009pycdt}
Freysoldt,~C.; Neugebauer,~J.; {Van De Walle},~C.~G. {Fully Ab Initio
  Finite-Size Corrections for Charged-Defect Supercell Calculations}.
  \emph{Phys. Rev. Lett.} \textbf{2009}, \emph{102}, 016402\relax
\mciteBstWouldAddEndPuncttrue
\mciteSetBstMidEndSepPunct{\mcitedefaultmidpunct}
{\mcitedefaultendpunct}{\mcitedefaultseppunct}\relax
\EndOfBibitem
\bibitem[Kumagai and Oba(2014)Kumagai, and Oba]{Kumagai2014Obapycdtcorrection}
Kumagai,~Y.; Oba,~F. {Electrostatics-based finite-size corrections for
  first-principles point defect calculations}. \emph{Phys. Rev. B}
  \textbf{2014}, \emph{89}, 195205\relax
\mciteBstWouldAddEndPuncttrue
\mciteSetBstMidEndSepPunct{\mcitedefaultmidpunct}
{\mcitedefaultendpunct}{\mcitedefaultseppunct}\relax
\EndOfBibitem
\bibitem[Adams(2001)]{adams2001relationship}
Adams,~S. {Relationship between bond valence and bond softness of alkali
  halides and chalcogenides.} \emph{Acta Crystallogr. Sect. B Struct. Sci.}
  \textbf{2001}, \emph{57}, 278--287\relax
\mciteBstWouldAddEndPuncttrue
\mciteSetBstMidEndSepPunct{\mcitedefaultmidpunct}
{\mcitedefaultendpunct}{\mcitedefaultseppunct}\relax
\EndOfBibitem
\bibitem[Sheppard \latin{et~al.}(2008)Sheppard, Terrell, and
  Henkelman]{Henkelman2008NEB}
Sheppard,~D.; Terrell,~R.; Henkelman,~G. {Optimization methods for finding
  minimum energy paths}. \emph{J. Chem. Phys.} \textbf{2008}, \emph{128},
  94107\relax
\mciteBstWouldAddEndPuncttrue
\mciteSetBstMidEndSepPunct{\mcitedefaultmidpunct}
{\mcitedefaultendpunct}{\mcitedefaultseppunct}\relax
\EndOfBibitem
\bibitem[Hoover(1985)]{Hoover1985thermostate}
Hoover,~W.~G. {Canonical dynamics: Equilibrium phase-space distributions}.
  \emph{Phys. Rev. A} \textbf{1985}, \emph{31}, 1695--1697\relax
\mciteBstWouldAddEndPuncttrue
\mciteSetBstMidEndSepPunct{\mcitedefaultmidpunct}
{\mcitedefaultendpunct}{\mcitedefaultseppunct}\relax
\EndOfBibitem
\bibitem[Wang \latin{et~al.}(2023)Wang, Panchal, and
  Pieremanuele]{Juefan2023MTPMDSE}
Wang,~J.; Panchal,~A.~A.; Pieremanuele,~C. {Strategies for fitting accurate
  machine-learned inter-atomic potentials for solid electrolytes
  machine-learned inter-atomic potentials for solid electrolytes}. \emph{Mater.
  Futur.} \textbf{2023}, \emph{2}, 015101\relax
\mciteBstWouldAddEndPuncttrue
\mciteSetBstMidEndSepPunct{\mcitedefaultmidpunct}
{\mcitedefaultendpunct}{\mcitedefaultseppunct}\relax
\EndOfBibitem
\bibitem[Zuo \latin{et~al.}(2020)Zuo, Chen, Li, Deng, Chen, Behler,
  Cs{\'{a}}nyi, Shapeev, Thompson, Wood, and Ong]{YunxingZuo2020MLIP}
Zuo,~Y.; Chen,~C.; Li,~X.; Deng,~Z.; Chen,~Y.; Behler,~J.; Cs{\'{a}}nyi,~G.;
  Shapeev,~A.~V.; Thompson,~A.~P.; Wood,~M.~A.; Ong,~S.~P. {Performance and
  Cost Assessment of Machine Learning Interatomic Potentials}. \emph{J. Phys.
  Chem. A} \textbf{2020}, \emph{124}, 731--745\relax
\mciteBstWouldAddEndPuncttrue
\mciteSetBstMidEndSepPunct{\mcitedefaultmidpunct}
{\mcitedefaultendpunct}{\mcitedefaultseppunct}\relax
\EndOfBibitem
\bibitem[Plimpton(1995)]{Steve1994LAMMPS}
Plimpton,~S. {Fast Parallel Algorithms for Short-Range Molecular Dynamics}.
  1995\relax
\mciteBstWouldAddEndPuncttrue
\mciteSetBstMidEndSepPunct{\mcitedefaultmidpunct}
{\mcitedefaultendpunct}{\mcitedefaultseppunct}\relax
\EndOfBibitem
\end{mcitethebibliography}

\begin{tocentry}
    \centering
    \includegraphics[scale=0.176]{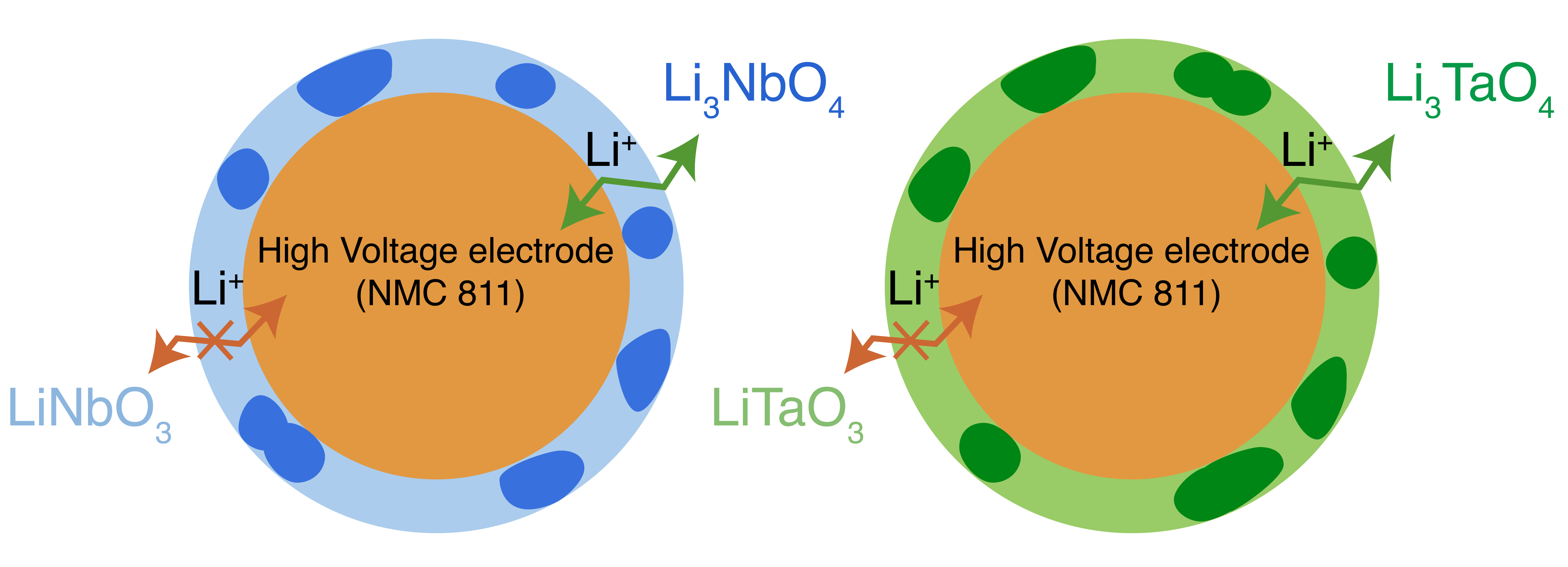}
\end{tocentry}

\end{document}